\documentclass[twocolumn]{aastex63}
\usepackage{booktabs}
\usepackage{amssymb}
\usepackage{lineno}
\revised{\today}
\shorttitle{GECKO Follow-up of S230518h}
\shortauthors{Paek et al.}
\graphicspath{{./}{figures/}}

\begin{document}


\title{GECKO Follow-up Observation of the Binary Neutron Star-Black Hole Merger Candidate S230518h}

\correspondingauthor{Myungshin Im}
\email{gregorypaek94@gmail.com, myungshin.im@gmail.com}
\author[0000-0002-6639-6533]{Gregory S.H. Paek}
\affiliation{SNU Astronomy Research Center, Astronomy Program, Department of Physics \& Astronomy, Seoul National University, 1 Gwanak-ro, Gwanak-gu, Seoul,  08826, Republic of Korea} 
\affiliation{Institute for Astronomy, University of Hawaii, 2680 Woodlawn Drive, Honolulu, HI 96822, USA}

\author[0000-0002-8537-6714]{Myungshin Im}
\affiliation{SNU Astronomy Research Center, Astronomy Program, Department of Physics \& Astronomy, Seoul National University, 1 Gwanak-ro, Gwanak-gu, Seoul,  08826, Republic of Korea} 

\author[0009-0003-1280-0099]{Mankeun Jeong}
\affiliation{SNU Astronomy Research Center, Astronomy Program, Department of Physics \& Astronomy, Seoul National University, 1 Gwanak-ro, Gwanak-gu, Seoul,  08826, Republic of Korea} 

\author[0000-0002-3118-8275]{Seo-Won Chang}
\affiliation{SNU Astronomy Research Center, Astronomy Program, Department of Physics \& Astronomy, Seoul National University, 1 Gwanak-ro, Gwanak-gu, Seoul,  08826, Republic of Korea}

\author[0009-0001-3706-5671]{Martin Moonkuk Hur}
\affiliation{SNU Astronomy Research Center, Astronomy Program, Department of Physics \& Astronomy, Seoul National University, 1 Gwanak-ro, Gwanak-gu, Seoul,  08826, Republic of Korea} 

\author[0009-0002-6238-2326]{YoungPyo Hong}
\affiliation{SNU Astronomy Research Center, Astronomy Program, Department of Physics \& Astronomy, Seoul National University, 1 Gwanak-ro, Gwanak-gu, Seoul,  08826, Republic of Korea} 

\author[0000-0002-0070-1582]{Sophia Kim}
\affiliation{SNU Astronomy Research Center, Astronomy Program, Department of Physics \& Astronomy, Seoul National University, 1 Gwanak-ro, Gwanak-gu, Seoul,  08826, Republic of Korea} 

\author[0009-0005-3910-0337]{Jaewon lee}
\affiliation{SNU Astronomy Research Center, Astronomy Program, Department of Physics \& Astronomy, Seoul National University, 1 Gwanak-ro, Gwanak-gu, Seoul,  08826, Republic of Korea} 

\author[0009-0003-0777-6717]{Dongjin Lee}
\affiliation{Pohang University of Science and Technology (POSTECH) \\ 
Pohang 37673, South Korea}

\author[0009-0003-3902-5678]{Seong-Heon Lee}
\affiliation{Pohang University of Science and Technology (POSTECH) \\ 
Pohang 37673, South Korea}

\author[0000-0001-7923-0674]{Jae-Hun Jung}
\affiliation{Pohang University of Science and Technology (POSTECH) \\ 
Pohang 37673, South Korea}

\author[0000-0002-1294-168X]{Joonho Kim}
\affiliation{Daegu National Science Museum, 20, Techno-daero 6-gil, Yuga-myeon, Dalseong-gun, Daegu 43023, Republic of Korea} 
\affiliation{SNU Astronomy Research Center, Astronomy Program, Department of Physics \& Astronomy, Seoul National University, 1 Gwanak-ro, Gwanak-gu, Seoul,  08826, Republic of Korea} 

\author[0000-0003-4412-7161]{Hyung Mok Lee}
\affiliation{SNU Astronomy Research Center, Astronomy Program, Department of Physics \& Astronomy, Seoul National University, 1 Gwanak-ro, Gwanak-gu, Seoul,  08826, Republic of Korea} 


\author[0000-0003-0043-3925]{Chung-Uk Lee}
\affiliation{Korea Astronomy and Space Science Institute, 776 Daedeokdae-ro, Yuseong-gu, Daejeon 34055, Korea}

\author[0000-0003-0562-5643]{Seung-Lee Kim}
\affiliation{Korea Astronomy and Space Science Institute, 776 Daedeokdae-ro, Yuseong-gu, Daejeon 34055, Korea}





\begin{abstract}

The gravitational wave (GW) event S230518h is a potential binary neutron star-black hole merger (NSBH) event that was detected during engineering run 15 (ER15), which served as the commissioning period before the LIGO-Virgo-KAGRA (LVK) O4a observing run. Despite its low probability of producing detectable electromagnetic emissions, we performed extensive follow-up observations of this event using the GECKO telescopes in the southern hemisphere. Our observation covered 61.7\% of the 90\% credible region, a $\rm 284\:deg^2$ area accessible from the southern hemisphere, reaching a median limiting magnitude of $R=21.6$ mag. In these images, we conducted a systematic search for an optical counterpart of this event by combining a CNN-based classifier and human verification. We identified 128 transient candidates, but no significant optical counterpart was found that could have caused the GW signal. Furthermore, we provide feasible KN properties that are consistent with the upper limits of observation. Although no optical counterpart was found, our result demonstrates both GECKO's efficient wide-field follow-up capabilities and usefulness for constraining properties of kilonovae from NSBH mergers at distances of $\sim 200$ Mpc.



\end{abstract}


\section{Introduction} \label{sec:intro}


The mergers of compact binary systems involving one or more neutron stars, such as binary neutron stars (BNS) or neutron star-black hole (NSBH) systems, represent promising multi-messenger events. These mergers emit gravitational waves (GWs) detectable by the GW detectors of the LIGO-Virgo-KAGRA Collaboration \citep{2020LRR....23....3A} and produce electromagnetic-wave (EM) counterparts powered by the radioactive decay of heavy elements synthesized in the ejecta, known as kilonovae (KNe; \citealt{2010MNRAS.406.2650M,2012ApJ...746...48M,2024arXiv240910651K}). NSBH mergers, in particular, are considered more promising for multi-messenger astronomy than BNS mergers, as suggested by recent studies \citep{2023PhRvD.107l4007G}, due to their louder GW signals and the potential to produce brighter KNe.



So far, the localization of GW sources has generally been poor, spanning hundreds to thousands of square degrees, which makes identifying the EM counterpart critical \citep{2020LRR....23....3A,2022ApJ...924...54P}. EM counterparts provide precise positions of GW events within a few arcseconds, facilitating the determination of their host galaxies \citep{2017ApJ...848L..12A}, environments \citep{2017ApJ...848L..28L,2017ApJ...849L..16I}, and redshifts \citep{2017ApJ...848L..28L}. The best and only example of such a KN associated with a GW source is AT2017gfo, a monumental event that opened the field of GW multi-messenger astronomy (MMA; \citealt{2017ApJ...848L..12A}). Through this event, researchers have studied both GW and EM signals to explore the environment of the GW source \citep{2017ApJ...849L..16I,2017ApJ...848L..28L,2018ApJ...859L...6L}, the chemical enrichment of heavy elements \citep{2017ApJ...848L..19C,2022Natur.612..223R,2024MNRAS.529.1154C}, and cosmology through the standard siren \citep{2017Natur.551...85A}. The work on AT2017gfo shows the great potential of MMA with KNe, one still needs more GW accompanying KNe like AT2017gfo to characterize the general KNe population through \citep{2017Sci...358.1559K,2017ApJ...851L..21V}, better understand the host galaxy properties \citep{2022ApJ...940...57N,2024ApJ...974..114J}, and reduce uncertainties in the Hubble constant derived from GW events as a way to resolve the Hubble tension \citep{2018Natur.562..545C,2019NatAs...3..940H,2019PhRvL.122f1105F}.


Despite improvements in GW detectors \citep{2020LRR....23....3A} and strategies and systems for optical follow-up observations since 2017, there has yet to be another EM counterpart detection for GW sources. Several main obstacles hinder the identification of the EM counterpart. Firstly, the dim and fast-decaying nature of KNe ($\rm \sim 1\:mag/day$; \citealt{2017Natur.551...64A,2017Natur.551...80K,2019MNRAS.489.5037B}). Secondly, even during the current O4b run of GW observations, the GW events are still poorly localized ($\rm >$ hundreds of $\rm deg^{2}$ for many events) due to the limited duty cycle and sensitivities of GW detectors \citep{2020LRR....23....3A}. Lastly, it is challenging and time-consuming to identify the KN associated with a GW event among an enormous number ($\mathcal{O}(10^3)$ to $\mathcal{O}(10^5)$) of other kinds of transients, KN contaminants, and image artifacts acting as false signals from extensive follow-up observations over the large localization area.

To overcome the above difficulties and efficiently find KNe associated with GW sources, we established the Gravitational-wave Electromagnetic Counterpart Korean Observatories (GECKO; \citealt{2020grbg.conf...25I,2024ApJ...960..113P}), a worldwide network of about 20 optical telescopes. These facilities enable continuous, round-the-clock follow-up observations across both the northern and southern hemispheres. GECKO includes several wide-field telescopes with an FOV greater than $\rm 1\:deg^2$, allowing extensive coverage of the wide localization areas typical of GW events. Additionally, the network contains narrow-field telescopes with aperture sizes ranging from 0.35 to 1.5 meters. During the past Advanced LIGO and Virgo O2 and O3 observing runs \citep{2019PhRvX...9c1040A}, the GECKO telescopes including the Korea Microlensing Telescope Network (KMTNet; \citealt{2016JKAS...49...37K}), were instrumental in the follow-up studies of GW sources \citep{2021ApJ...916...47K}, including the famous GW170817 \citep{2017ApJ...848L..12A,2017Natur.551...71T} and GW190425 \citep{2024ApJ...960..113P}.

We report the results of our follow-up observation of the GW event S230518h, which occurred during the engineering run of the O4 on 2023-05-18 (UT), just before the official start of the O4 run. Classified as a likely NSBH merger event, S230518h presented a unique opportunity to test and refine the GECKO follow-up process. Its 90\% credible region, spanning 460 deg$^2$, was manageable with the GECKO facility, enabling an intensive observation campaign using three 1.6 m telescopes of the KMTNet and two small telescopes, RASA36 and LSGT, from the GECKO network. Although the probability of detecting an EM counterpart, as estimated from GW signals, was low, comprehensive observations of this event provided the opportunity to refine constraints on its expected properties and potentially discover it, thereby challenging the theoretical predictions derived from GW signals. We identified transient candidates during this campaign, evaluated them through rigorous screening procedures, and applied upgraded data analysis techniques, including an efficient alert system, an automatic data reduction pipeline, and a Machine Learning (ML)-based transient search procedure. Additionally, the depth of our observations allowed us to place meaningful constraints on KN properties, such as ejecta mass and composition, advancing our understanding of NSBH merger physics. This study underscores the importance of coordinated follow-up campaigns in refining observational strategies and enhancing the scientific yield of multi-messenger astronomy, even in the absence of a confirmed counterpart. Throughout this paper, we adopt the AB magnitude system.

\section{Follow-up Observation of S230518h}\label{sec:s230518h}

\subsection{Properties of S230518h}\label{ss:gwevent}

On 2023 May 18 at 12:59:26 (UT), six days before the official start of the O4 run on May 24, the GW event designated as S230518h was detected by the LIGO Hanford Observatory and LIGO Livingston Observatory of the LVK Collaboration with a false alarm rate of one in 98 years \citep{2023GCN.33813....1L}. This event was reported to have at least one neutron star in the binary merging system (\texttt{HasNS} $>$ 99\%), and the classification according to the \texttt{PyCBCLive} \citep{2021ApJ...923..254D}, \texttt{GstLAL} \citep{2019arXiv190108580S}, and \texttt{MBTAOnline} \citep{2021CQGra..38i5004A} pipelines is that S230518h is an NSBH event with probability of 86\% \citep{2023GCN.33813....1L}. After the preliminary alert was issued,  the initial localization, based on Bayestar, was released at 13:26 UTC on the same day, and the update using Bilby was issued at 15:33 UTC. The GW-based luminosity distance and projected 2D localization area were refined from the initial Bayestar value to the updated Bilby offline analysis as follows: the luminosity distance from $276 \pm 79$ Mpc to $204 \pm 57$ Mpc, and the 90\% credible region (CR90) $1,002 \:\mathrm{deg}^2$ to $460 \:\deg^2$ \citep{2023GCN.33813....1L,2023GCN.33816....1L,2023GCN.33884....1L}. Due to the timing of the update, the follow-up observations were based on the initial localization, while the post-analysis utilized the updated localization. The CR90 area had an elongated shape spanning both the northern and southern hemispheres, but it was more slanted towards the southern hemisphere. Unfortunately, the Sun’s position (RA = 73.25 deg, Dec = 21.48 deg) created an unobservable gap in the northern hemisphere, preventing northern facilities from conducting follow-up observations \citep{2024PASP..136k4201A}. Furthermore, the probability of having a remnant that could potentially produce a detectable EM signal was less than 1\%. Despite these challenges, we initiated GECKO follow-up observations to provide useful constraints on the NSBH merger event (Figure \ref{fig:observation}).

\subsection{GECKO Follow-up Strategy and Facilities}\label{ss:facility}

The GECKO follow-up observations were conducted using five telescopes from three facilities, each playing a complementary role in the campaign. The three 1.6 m telescopes of KMTNet \citep{2016JKAS...49...37K}, located in Australia, South Africa, and Chile, provided continuous round-the-clock coverage of the event. Each KMTNet telescope has an FOV of 2 $\times$ 2 deg$^2$ and is equipped with $BVRI$ filters. Observations for this campaign were conducted primarily in the $R$-band, with each field observed using a 2-point dithering pattern to fill CCD gaps and enhance depth.

The KMTNet observations were aligned with the KMTNet Synoptic Survey of Southern Sky (KS4; Im et al. in preparation), which has been conducted since 2018 to prepare deep reference images for transient search. KS4 provides $BVRI$ reference images for the Southern Hemisphere, focusing on areas where deep imaging data are scarce. Using the predefined KS4 tiling pattern ensures compatibility between the newly observed images and the reference frames, minimizing subtraction artifacts and enabling efficient image differencing for transient identification.

RASA36, located in Chile, has a $0.36\:\mathrm{m}$ aperture size with a large FOV of $\sim 2.67 \times 2.67 \:\mathrm{deg}^2$. Its observations complemented those of KMTNet by targeting areas near the South Celestial Pole, where KMTNet's declination limits prevented coverage. RASA36 was equipped with a single $r$-band filter, which was used for all observations. To streamline the observation process, RASA36 adhered to the KS4 tiling pattern developed for KMTNet, leveraging its larger FOV to maximize coverage efficiency while maintaining compatibility with the KS4 reference frame.

The Lee Sang Gak Telescope (LSGT; \citealt{2015JKAS...48..207I}), located in Australia, provided targeted observations of individual host galaxies. The $0.43\:\mathrm{m}$ telescope is equipped with the SNUCAM-II camera, capable of imaging in $ugriz$ broad-band filters as well as a suite of medium-band filters \citep{2017JKAS...50...71C}. For this campaign, LSGT utilized the $r$-band to observe prioritized host galaxy candidates identified through the \texttt{GLADE+} catalog \citep{2022MNRAS.514.1403D} and scored based on their $K$-band luminosity.

The complementary roles of these telescopes ensured efficient coverage of the GW localization area and host galaxies. KMTNet and RASA36 prioritized wide-field tiling observations to cover the CR90 area, while LSGT focused on high-probability host galaxies. This strategic combination enabled the GECKO network to maximize its observational efficiency and scientific return, even with the constraints imposed by the localization shape and the Sun's position.

\subsection{Observation Campaign}\label{ss:observation}
\begin{figure*}
    \centering
    \includegraphics[width=1\linewidth]{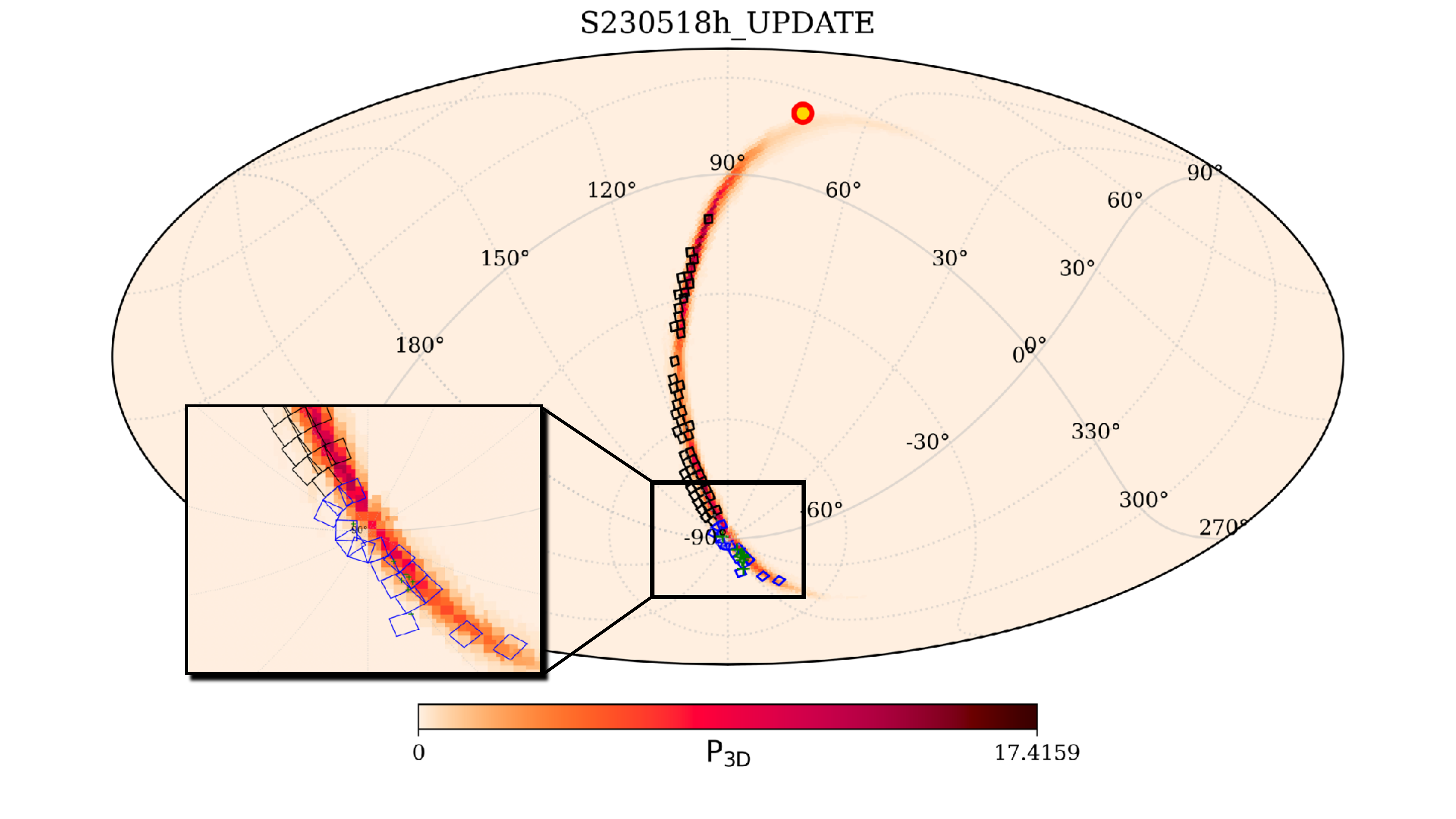}
    \caption{Coverage map of the KMTNet and RASA36 tiling observations, and the LSGT galaxy-targeted observation for S230518h. The color map represents the \texttt{update} localization probability map. Black boxes denote the FOV of KMTNet, blue boxes denote the FOV of RASA36, and green crosses mark the LSGT pointings. The red circle indicates the position of the Sun on May 18, 2023 (UTC).}
    \label{fig:observation}
\end{figure*}

We used the \texttt{GeckoDigestor} \citep{geckodigestor}, our real-time alert software, to automatically process the GW alert and generate a target list. This process selects and prioritizes host galaxy candidates within a GW localized volume, and creates the observation sequences that are based on wide-field tiling observation pattern to cover the whole CR90 and galaxy-target observation to observe galaxies with a high probability of hosting an NSBH merger in sequence \citep{2021ApJ...916...47K,2024ApJ...960..113P}. The \texttt{GLADE+} catalog was used as the input catalog of galaxies \citep{2022MNRAS.514.1403D}. The prioritization involved scoring host galaxy candidates primarily using their $K$-band luminosity as a proxy for stellar mass, which correlates with the CBC rate based on simulation results. For objects without $K$-band luminosity, we used $B$-band instead. For such cases, the highest rank of the $B$-band only galaxies is given as the rank just below the lowest rank of galaxies with $K$-band. See \citet{2024ApJ...960..113P} for details on the prioritization scheme. The number of prioritized host galaxy candidates is 13,741 for the Bilby (\texttt{update}) localization volume and 18,500 for the Bayestar (\texttt{preliminary}) localization volume. For the tiling observation, each tile of both KMTNet and RASA36 was prioritized based on the summed scores of host galaxy candidates. RASA36 covered areas that were missed by KMTNet, in particular the area near the South Celestial Pole where the declination limit of KMTNet prevented its observations (Figure \ref{fig:observation}).



The observations were carried out using the $R$-band for KMTNet with an on-source exposure time of 120 seconds per frame, typically using a 2-point dithering pattern (120 seconds × 2) to cover gaps in the CCD chips and increase sensitivity (resulting in 240 seconds per field). However, in some cases, observations consisted of only a single 120-second frame or up to three 120-second frames per field. For RASA36, $r$-band data were taken with exposure times ranging from 420 seconds to 4,380 seconds, where each frame had an exposure time of 60 seconds. For LSGT, $r$-band observations were conducted with a fixed exposure time of 180 seconds per frame, with total exposure times ranging from 180 seconds (1 frame) to 5,040 seconds (28 frames) for each of the 12 galaxy targets observed.


The first data were obtained approximately 5 hours after the GW trigger using KMTNet-SAAO. The delay was primarily due to the visibility constraints at the observing site. Following the initial observations at KMTNet-SAAO, the survey continued with KMTNet telescopes at CTIO (Chile) and SSO (Australia), along with RASA36, to cover the remaining tiles. The field coverage followed a predefined set of field locations (tiles) that were used for the KS4 survey. RASA36 adhered to the KS4 tiling pattern, despite having a larger FOV than KMTNet, to utilize the KS4 reference frame and simplify the observation process. LSGT independently targeted the sorted host galaxy candidates one by one. Observations continued until 1.3 days after the GW trigger. Each observation is summarized in Table \ref{tab:observation}.


As a result, we managed to cover 71 tiles using KMTNet (55 tiles, 77.5\% of the total observed area) and RASA36 (16 tiles), along with four host galaxy candidates targeted by LSGT within the \texttt{update} localization volume. These observations spanned approximately 284 deg$^2$ ($\sim$61.7\% of the CR90 area), covering 6,461 host galaxy candidates ($\sim$47.0\%) and accounting for 52.3\% of the total score. The coverage included 72 deg$^2$ (55.1\%) of the 50\% confidence area and 220 deg$^2$ (47.8\%) of the 90\% confidence area. Observations reached limiting magnitudes of up to 23.29 mag for KMTNet, 19.67 mag for RASA36, and 20.80 mag for LSGT. The locations of the covered fields and their depths are summarized in Table \ref{tab:observation}.

\begin{longtable*}{ccccccccc}
    \caption{GECKO Observation of S230518h} \\
    \toprule
    \midrule
    Observatory & Filter & $\Delta$t & Date-obs & Depth   & Exposure & Galaxies or & RA      & Dec   \\
                &        &           &          &         & Time     & GLADE Name  &         &       \\
                &        & (days)    & (UTC)    & (ABmag) & (s)      &             & (hms)   & (dms) \\
    \hline
    \endfirsthead

    \caption[]{GECKO Observation of S230518h (continued)} \\
    \toprule
    \midrule
    Observatory & Filter & $\Delta$t & Date-obs & Depth   & Exposure & Galaxies or & RA      & Dec   \\
                &        &           &          &         & Time     & GLADE Name  &         &       \\
                &        & (days)    & (UTC)    & (ABmag) & (s)      &             & (hms)   & (dms) \\
    \hline
    \endhead

    \hline
    \multicolumn{9}{p{\textwidth}}{\footnotesize \textbf{Note:} Galaxy names are listed for LSGT observations, as it was used for galaxy-targeted observations with its narrow FOV. An asterisk (*) indicates galaxies matched with the \texttt{update} localization volume, while others are matched with the \texttt{initial} localization volume. For non-LSGT observations, the number of galaxies within the observed region is provided.} \\
    
    \endfoot

    \hline
    \endlastfoot

    KMTNet SAAO & $R$ & 0.1868 & 2023-05-18T17:28:27 & 21.56 & 120 & 127 & 07:46:28 & -68:00:00 \\
    KMTNet SAAO & $R$ & 0.1901 & 2023-05-18T17:33:10 & 22.16 & 240 & 248 & 07:07:21 & -32:00:00 \\
    KMTNet SAAO & $R$ & 0.1944 & 2023-05-18T17:39:18 & 21.56 & 240 & 227 & 07:10:00 & -37:59:59 \\
    KMTNet SAAO & $R$ & 0.1990 & 2023-05-18T17:46:03 & 22.09 & 240 & 183 & 07:08:06 & -35:59:59 \\
    KMTNet SAAO & $R$ & 0.2039 & 2023-05-18T17:53:06 & 22.14 & 240 & 107 & 08:11:42 & -77:59:59 \\
    KMTNet SAAO & $R$ & 0.2083 & 2023-05-18T17:59:21 & 22.09 & 240 & 73 & 08:13:42 & -79:59:59 \\
    KMTNet SAAO & $R$ & 0.2125 & 2023-05-18T18:05:23 & 21.68 & 240 & 156 & 07:43:43 & -71:59:59 \\
    KMTNet SAAO & $R$ & 0.2210 & 2023-05-18T18:17:40 & 22.24 & 240 & 126 & 07:45:13 & -69:59:59 \\
    KMTNet SAAO & $R$ & 0.2243 & 2023-05-18T18:22:24 & 23.29 & 120 & 39 & 08:16:33 & -82:00:00 \\
    KMTNet SAAO & $R$ & 0.2298 & 2023-05-18T18:30:18 & 21.48 & 240 & 100 & 08:10:12 & -76:00:00 \\
    KMTNet SAAO & $R$ & 0.2342 & 2023-05-18T18:36:38 & 22.61 & 240 & 24 & 08:06:45 & -67:59:59 \\
    KMTNet SAAO & $R$ & 0.2385 & 2023-05-18T18:42:51 & 22.08 & 240 & 106 & 07:39:34 & -75:59:59 \\
    KMTNet SAAO & $R$ & 0.2418 & 2023-05-18T18:47:35 & 21.72 & 120 & 63 & 08:43:38 & -83:59:59 \\
    
    KMTNet CTIO & $R$ & 0.4369 & 2023-05-18T23:28:36 & 21.93 & 480 & 262 & 07:09:08 & -33:59:59 \\
    KMTNet CTIO & $R$ & 0.4502 & 2023-05-18T23:47:45 & 21.65 & 240 & 195 & 07:34:44 & -52:00:00 \\
    KMTNet CTIO & $R$ & 0.4545 & 2023-05-18T23:53:58 & 21.97 & 240 & 87 & 07:42:23 & -54:00:00 \\
    KMTNet CTIO & $R$ & 0.4594 & 2023-05-19T00:00:56 & 21.95 & 240 & 85 & 07:45:27 & -58:00:00 \\
    KMTNet CTIO & $R$ & 0.4638 & 2023-05-19T00:07:19 & 20.11 & 240 & 104 & 07:36:35 & -77:59:59 \\
    KMTNet CTIO & $R$ & 0.4681 & 2023-05-19T00:13:28 & 21.14 & 240 & 178 & 07:36:55 & -56:00:00 \\
    KMTNet CTIO & $R$ & 0.4745 & 2023-05-19T00:22:42 & 21.31 & 240 & 53 & 08:08:08 & -71:59:59 \\
    KMTNet CTIO & $R$ & 0.4820 & 2023-05-19T00:33:29 & 21.38 & 360 & 119 & 07:44:31 & -60:00:00 \\
    KMTNet CTIO & $R$ & 0.4958 & 2023-05-19T00:53:26 & 18.78 & 120 & 55 & 08:09:03 & -74:00:00 \\
    KMTNet CTIO & $R$ & 0.4992 & 2023-05-19T00:58:16 & 21.31 & 240 & 51 & 07:36:35 & -48:00:00 \\
    KMTNet CTIO & $R$ & 0.5035 & 2023-05-19T01:04:32 & 20.47 & 240 & 34 & 08:07:23 & -70:00:00 \\
    
    RASA36 & $r$ & 0.5072 & 2023-05-19T01:09:51 & 18.79 & 420 & 180 & 07:30:00 & -86:00:00 \\

    KMTNet CTIO & $R$ & 0.5084 & 2023-05-19T01:11:36 & 21.21 & 240 & 80 & 07:47:31 & -66:00:00 \\
    KMTNet CTIO & $R$ & 0.5133 & 2023-05-19T01:18:34 & 21.44 & 240 & 0 & 08:46:49 & -78:00:00 \\
    KMTNet CTIO & $R$ & 0.5177 & 2023-05-19T01:24:51 & 20.58 & 240 & 0 & 08:40:50 & -76:00:01 \\
    KMTNet CTIO & $R$ & 0.5241 & 2023-05-19T01:34:09 & 20.48 & 240 & 1 & 08:36:13 & -73:59:59 \\
    KMTNet CTIO & $R$ & 0.5273 & 2023-05-19T01:38:45 & 21.80 & 240 & 0 & 08:32:32 & -72:00:00 \\
    
    RASA36 & $r$ & 0.5354 & 2023-05-19T01:50:26 & 19.17 & 1860 & 77 & 09:00:00 & -86:00:00 \\
    RASA36 & $r$ & 0.5596 & 2023-05-19T02:25:18 & 18.65 & 960 & 72 & 12:00:00 & -88:00:00 \\
    RASA36 & $r$ & 0.5799 & 2023-05-19T02:54:31 & 17.58 & 780 & 33 & 14:24:00 & -88:00:00 \\
    RASA36 & $r$ & 0.6086 & 2023-05-19T03:35:53 & 17.70 & 540 & 10 & 10:30:00 & -86:00:00 \\
    RASA36 & $r$ & 0.6306 & 2023-05-19T04:07:26 & 17.52 & 600 & 0 & 19:12:00 & -80:00:00 \\
    RASA36 & $r$ & 0.6539 & 2023-05-19T04:41:07 & 18.81 & 1080 & 124 & 16:48:00 & -88:00:00 \\
    RASA36 & $r$ & 0.6758 & 2023-05-19T05:12:39 & 18.14 & 420 & 101 & 19:38:10 & -84:00:00 \\
    RASA36 & $r$ & 0.7056 & 2023-05-19T05:55:26 & 19.63 & 1500 & 84 & 20:41:22 & -82:00:00 \\
    RASA36 & $r$ & 0.7468 & 2023-05-19T06:54:49 & 18.25 & 600 & 294 & 19:12:00 & -88:00:00 \\
    RASA36 & $r$ & 0.7781 & 2023-05-19T07:39:54 & 18.28 & 660 & 126 & 21:00:00 & -86:00:00 \\
    RASA36 & $r$ & 0.8022 & 2023-05-19T08:14:40 & 19.08 & 1020 & 109 & 20:25:31 & -76:00:00 \\
    RASA36 & $r$ & 0.8256 & 2023-05-19T08:48:16 & 18.67 & 1080 & 101 & 19:30:00 & -86:00:00 \\
    
    KMTNet SSO & $R$ & 0.8377 & 2023-05-19T09:05:46 & 21.47 & 240 & 97 & 07:17:50 & -35:59:59 \\
    KMTNet SSO & $R$ & 0.8434 & 2023-05-19T09:13:55 & 21.24 & 120 & 77 & 07:29:01 & -60:00:00 \\
    KMTNet SSO & $R$ & 0.8469 & 2023-05-19T09:19:01 & 21.53 & 240 & 113 & 07:41:53 & -73:59:59 \\
    KMTNet SSO & $R$ & 0.8503 & 2023-05-19T09:23:49 & 21.25 & 120 & 40 & 07:54:32 & -61:59:59 \\
    
    RASA36 & $r$ & 0.8506 & 2023-05-19T09:24:14 & 19.27 & 1500 & 63 & 20:44:44 & -72:00:00 \\
    
    KMTNet SSO & $R$ & 0.8556 & 2023-05-19T09:31:32 & 21.66 & 240 & 9 & 08:06:14 & -65:59:59 \\
    KMTNet SSO & $R$ & 0.8590 & 2023-05-19T09:36:25 & 21.33 & 120 & 0 & 08:29:32 & -69:59:59 \\
    
    RASA36 & $r$ & 0.8747 & 2023-05-19T09:59:01 & 17.91 & 1200 & 47 & 19:51:43 & -82:00:00 \\

    LSGT & $r$ & 0.9453 & 2023-05-19T11:40:25 & 20.60 & 2700 & \texttt{1527565} & 10:14:46 & -88:27:41 \\
    LSGT & $r$ & 0.9796 & 2023-05-19T12:29:45 & 20.67 & 2700 & \texttt{1208642} & 19:35:05 & -80:45:07 \\
    LSGT & $r$ & 1.0136 & 2023-05-19T13:18:42 & 20.71 & 2700 & \texttt{1208719} & 20:03:06 & -83:05:48 \\
    LSGT & $r$ & 1.0476 & 2023-05-19T14:07:39 & 20.76 & 2700 & \texttt{1157263}* & 20:31:46 & -84:03:30 \\ 
    LSGT & $r$ & 1.0925 & 2023-05-19T15:12:18 & 20.80 & 2700 & \texttt{1167888}* & 20:30:36 & -84:29:25 \\ 
    LSGT & $r$ & 1.1290 & 2023-05-19T16:04:58 & 20.68 & 2700 & \texttt{1208712} & 20:30:12 & -83:25:35 \\
    LSGT & $r$ & 1.1630 & 2023-05-19T16:53:52 & 20.73 & 5040 & \texttt{1163100}* & 20:04:28 & -84:11:24 \\ 
    
    KMTNet SAAO & $R$ & 1.1746 & 2023-05-19T17:10:55 & 21.64 & 240 & 36 & 06:44:29 & -19:08:56 \\
    KMTNet SAAO & $R$ & 1.1752 & 2023-05-19T17:11:47 & 21.77 & 360 & 409 & 06:50:13 & -24:53:37 \\
    KMTNet SAAO & $R$ & 1.1778 & 2023-05-19T17:15:29 & 21.17 & 120 & 47 & 06:21:24 & -11:29:21 \\
    KMTNet SAAO & $R$ & 1.1887 & 2023-05-19T17:31:12 & 21.68 & 240 & 366 & 06:45:31 & -22:58:43 \\
    KMTNet SAAO & $R$ & 1.1930 & 2023-05-19T17:37:22 & 21.71 & 240 & 341 & 06:48:59 & -26:48:30 \\
    KMTNet SAAO & $R$ & 1.1975 & 2023-05-19T17:43:46 & 21.70 & 240 & 348 & 06:57:30 & -26:48:30 \\
    KMTNet SAAO & $R$ & 1.2019 & 2023-05-19T17:50:06 & 21.74 & 240 & 197 & 06:40:54 & -21:03:49 \\
    KMTNet SAAO & $R$ & 1.2055 & 2023-05-19T17:55:17 & 21.42 & 120 & 218 & 06:56:23 & -28:43:24 \\
    KMTNet SAAO & $R$ & 1.2194 & 2023-05-19T18:15:24 & 21.48 & 120 & 224 & 06:59:14 & -29:59:59 \\
    KMTNet SAAO & $R$ & 1.2220 & 2023-05-19T18:19:11 & 21.53 & 120 & 59 & 06:58:36 & -24:53:37 \\
    KMTNet SAAO & $R$ & 1.2257 & 2023-05-19T18:24:24 & 21.38 & 120 & 93 & 07:05:03 & -28:43:24 \\
    
    LSGT & $r$ & 1.2278 & 2023-05-19T18:27:14 & 19.01 & 180 & \texttt{1213035} & 20:30:02 & -86:11:15 \\
    
    KMTNet SAAO & $R$ & 1.2305 & 2023-05-19T18:31:24 & 21.85 & 240 & 100 & 07:26:54 & -81:59:59 \\
    KMTNet SAAO & $R$ & 1.2341 & 2023-05-19T18:36:36 & 21.58 & 120 & 102 & 07:27:16 & -44:00:00 \\
    
    LSGT & $r$ & 1.2360 & 2023-05-19T18:39:02 & 18.82 & 180 & \texttt{1163100} & 20:04:28 & -84:11:24 \\
    
    KMTNet SAAO & $R$ & 1.2394 & 2023-05-19T18:44:09 & 21.76 & 240 & 163 & 07:27:43 & -50:00:00 \\
    
    LSGT & $r$ & 1.2413 & 2023-05-19T18:46:39 & 19.43 & 360 & \texttt{1335592}* & 20:31:45 & -84:03:30 \\ 
    
    KMTNet SAAO & $R$ & 1.2456 & 2023-05-19T18:53:05 & 21.88 & 240 & 71 & 07:39:49 & -50:00:00 \\
    
    LSGT & $r$ & 1.2470 & 2023-05-19T18:54:49 & 19.52 & 360 & \texttt{1335592} & 20:31:45 & -84:03:30 \\
    
    KMTNet SAAO & $R$ & 1.2488 & 2023-05-19T18:57:44 & 21.62 & 120 & 85 & 07:38:10 & -62:00:00 \\
    
    LSGT & $r$ & 1.2515 & 2023-05-19T19:01:19 & 20.18 & 1260 & \texttt{1335592} & 20:31:45 & -84:03:30 \\
    
    KMTNet SAAO & $R$ & 1.2531 & 2023-05-19T19:03:55 & 21.81 & 240 & 28 & 07:50:46 & -56:00:00 \\
    KMTNet SAAO & $R$ & 1.2639 & 2023-05-19T19:19:23 & 20.57 & 240 & 2 & 08:00:00 & -59:59:59 \\
    KMTNet SAAO & $R$ & 1.2689 & 2023-05-19T19:26:36 & 19.63 & 240 & 4 & 08:54:51 & -80:00:00 \\
    KMTNet SAAO & $R$ & 1.2732 & 2023-05-19T19:32:50 & 21.65 & 240 & 8 & 09:06:12 & -81:59:59 \\
    
    RASA36 & $r$ & 1.7538 & 2023-05-20T07:04:57 & 20.09 & 4380 & 150 & 20:43:38 & -84:00:00

\label{tab:observation}
\end{longtable*}

\section{Data Reduction and Analysis}\label{sec:analysis}

We used two different pipelines to reduce the data. The \texttt{gpPy} \citep{gppy} software reduced the data taken from RASA36 and LSGT. This software is initially aimed to automatically process the daily data for the Intensive Monitoring Survey of Nearby Galaxies (IMSNG; \citealt{2015JKAS...48..207I}) from various facilities to monitor young supernovae (SNe) or other transients. For the KMTNet data, the data reduction pipeline from the KMTNet operation team performed the reduction process up to flat-fielding, cross-talk correction and handed over the reduced data to us. Then, we used our in-house pipeline to refine the astrometry and photometry solutions and produced the source catalogs (Jeong et al. in preparation). Here, we describe the overall process for each pipeline.


\subsection{\texttt{gpPy}, Data Reduction Pipeline}\label{ss:gppy}
\texttt{gpPy} is an automatic pipeline that covers basic data reduction, astrometry, image stacking, calibration, photometry, image subtraction, and transient search using multi-threading on CPUs. \texttt{gpPy} standardized data from various telescope and camera configurations into a uniform format, which is then stored in our database. The data from RASA36 and LSGT were processed through \texttt{gpPy}, and the basic processing tasks are detailed in \citet{2024ApJ...960..113P}. The reference frames in the $r$-band for image subtraction were sourced from Pan-STARRS DR1 \citep{2016arXiv161205560C} and KS4. While there is a difference between the $r$- and $R$-bands, their central wavelengths (622 nm for $r$-band and 658 nm for $R$-band) and wavelength ranges (541–698 nm for $r$-band and 550–750 nm for $R$-band) are sufficiently similar. As a result, although some residuals may persist in the subtracted image, it remains effective for removing the underlying galaxy and isolating the newly emerged transient.


\subsection{KS4 Pipeline}\label{ss:ks4}
The KS4 pipeline involves quality assurance, bad pixel flagging and mask generation, and astrometric calibration and photometric calibration designed for the KMTNet imaging data taken for the KS4 survey. The pipeline employs rigorous quality checks, such as filtering out images with bad seeing conditions or tracking errors. The KS4 pipeline subdivided the KMTNet images and produced images with an astrometry accuracy of about $0\farcs3$ root mean square. Importantly, the pipeline re-derives the photometry zero points by finding the photometry solution over a 2-dimensional image area and making necessary corrections to the images based on this solution. Further details of the pipeline will be provided in Jeong et al. (in preparation).

Similar to \texttt{gpPy}, the KS4 pipeline also includes image subtraction and transient search when provided with reference images. At the time of S230518h observation, we had KS4 reference images for 45 KMTNet tiles among the 55 tiles. For the 10 tiles without the KS4 images, we combined the $r$-band PS1 images to create mosaic reference images for differential image analysis (DIA). The transient search algorithm in the KS4 and \texttt{gpPy} pipelines is described in the next section.
\section{Transient Search and Analysis}\label{sec:search}

\subsection{Image Subtraction and Filtering}\label{ss:subtraction}


After the data processing as described in Section \ref{ss:ks4}, The image subtraction process began with running \texttt{SExtractor} \citep{1996A&AS..117..393B} on both science and reference images to measure their seeing conditions and background characteristics. Then, we performed the reference image subtraction using \texttt{HOTPANTS} \citep{2015ascl.soft04004B}. This process includes convolving the images (reference or science) with better seeing to match the images with worse seeing, matching the imaging flux scaling of the science and reference images, and creating the subtracted (difference) image. Then, we ran \texttt{SExtractor} again to detect sources remaining in the subtracted images. 

During DIA, many artificial sources (bogus) can appear in the subtracted image due to imperfect image subtraction. To filter out transients from bogus signals, we follow the following steps: 

\begin{enumerate}

    \item Filtering using the \texttt{SExtractor} parameters

    \item Using Real/Bogus classifier

    \item Visual inspection

\end{enumerate}

First, we set the criteria of \texttt{SExtractor} parameters to reject the sources with extended profiles or other non-PSF-like characteristics and we aimed at detecting sources with $\rm SNR> 5$. 

Our initial search yielded about 13 million candidates. This large number includes various types of “false detections” arising from the subtraction process, along with real transients and moving objects such as asteroids. To systematically remove these false detections, we applied the following filtering criteria:

\begin{enumerate}
    \item \textbf{Elongated Sources (\texttt{ELLIPTICITY})}:\\
    To remove excessively elongated sources, sources with ellipticity greater than four times the median ellipticity of all sources in the image are excluded.
    
    
    
    \item \textbf{Inverted Subtraction Residuals}:\\
        To mitigate subtraction artifacts, we inverted the subtracted image by multiplying all pixel values by $-1$, effectively reversing the sign of the pixel values. Sources detected in the inverted image that match within half the median seeing value of point sources, and are brighter by more than 0.1 mag compared to the original subtraction, are excluded. This process helps to identify and remove paired dark and bright spots that often arise as subtraction artifacts.
    
    \item \textbf{Inverted Reference Image Defects}:\\
        To address defects in the reference image that produce spurious detections, we inverted the reference image and repeated the source detection process. Sources detected in the inverted reference image that match within half the median seeing value of point sources, and are brighter than 0.5 mag, are excluded. This approach removes artifacts caused by negative pixel values in the reference image, which can mimic transient candidates when subtracted from the science image.
    
    \item \textbf{Low Signal-to-Noise Ratio (SNR) Filter}:\\
        Sources with an SNR less than 5 are excluded to filter out low-confidence detections.
    
    \item \textbf{Unusual Source Profile (FWHM)}:\\
        Sources with full-width at half-maximum (FWHM) less than 0.7 or greater than 2.4 times the median FWHM are excluded to remove sources with atypical profiles.
    
    \item \textbf{Cosmic Rays and Coverage Flags (\texttt{FLAGS})}:\\
        Sources flagged with \texttt{SExtractor}'s \texttt{FLAGS} $\geq 4$ or a mask flag $\geq 1$ are excluded due to contamination from cosmic rays or inadequate coverage in the reference image.
    
    \item \textbf{Dipole Pattern Detection}:\\
        Sources exhibiting a dipole pattern, likely caused by misaligned subtraction, are excluded.
    
    \item \textbf{Anomalously High Background}:\\
        Sources with a background level more than 2$\sigma$ above the median are excluded to avoid detections in noisy regions.
    
    \item \textbf{Edge-of-Image Sources}:\\
        Sources located within the outermost 1\% of pixels, susceptible to edge effects, are excluded.
    
    \item \textbf{Null Pixel Sources}:\\
        Sources containing null pixels from failed subtraction calculations are excluded to ensure data integrity.
    
    \item \textbf{Exclusion of Solar System Objects}:\\
        Sources matched within 5 arcseconds to known solar system objects (e.g., asteroids or comets) are excluded using \texttt{SkyBot} \citep{2006ASPC..351..367B}.
\end{enumerate}







\begin{longtable*}{ccccc}
    \caption{Rejection Statistics by Flag} \\
    \toprule
    \midrule
    \textbf{Flag} & \textbf{Description} & \textbf{Criteria} & \textbf{Rejected Sources} & \textbf{Percentage (\%)} \\
    \hline
    \endfirsthead

    \caption[]{Rejection Statistics by Flag (continued)} \\
    \toprule
    \midrule
    \textbf{Flag} & \textbf{Description} & \textbf{Criteria} & \textbf{Rejected Sources} & \textbf{Percentage (\%)} \\
    \hline
    \endhead

    \hline
    \endfoot

    \hline
    \endlastfoot

    \texttt{flag\_0}  & Exclusion of Solar System Objects   & $\Delta \theta \leq 5^{\prime\prime}$ from known solar system objects & 7          & 0.0\%  \\
    \texttt{flag\_1}  & Inverted Subtraction Residuals      & $\Delta m > 0.1$ mag in inverted subtraction                          & 7,084,510  & 55.3\% \\
    \texttt{flag\_2}  & Inverted Reference Image Defects    & $\Delta m > 0.5$ mag in inverted reference image                      & 6,349,596  & 49.6\% \\
    \texttt{flag\_3}  & Edge-of-Image Sources               & Edge region: outermost $1\%$ pixels                                   & 233,249    & 1.8\%  \\
    \texttt{flag\_5}  & Elongated Sources                   & $\mathrm{Ellipticity} > 4\times \mathrm{median}$                      & 10,429,269 & 81.4\% \\
    \texttt{flag\_6}  & Cosmic Rays and Coverage Flags      & $\texttt{FLAGS} \geq 4$ or $\texttt{MaskFlag} \geq 1$                 & 1,627,697  & 12.7\% \\
    \texttt{flag\_7}  & Unusual Source Profile              & $\mathrm{FWHM} < 0.7 \times \mathrm{median}$ or $> 2.4 \times \mathrm{median}$ & 3,686,852  & 28.8\% \\
    \texttt{flag\_8}  & Anomalously High Background         & $\mathrm{Background} > 2\sigma$ above median                          & 496,673    & 3.9\%  \\
    \texttt{flag\_9}  & Low SNR                             & $\mathrm{SNR} < 5$                                                    & 4,884,083  & 38.1\% \\
    \texttt{flag\_10} & Dipole Pattern Detection            & Dipole pattern in subtraction image                                   & 1,266,075  & 9.9\%  \\
    \texttt{flag\_11} & Null Pixel Sources                  & Sources with null pixel values                                        & 32,997     & 0.3\%  \\
    \hline
    Total             & Filtered Sources                    &                                                                       & 12,806,361 & 96.41\%  

\label{tab:flag}
\end{longtable*}

After filtering, the remaining sources were reduced to 476,196 in total, with 297,558 sources originating from subtracted images using the KS4 reference frames and 178,636 sources using the PS1 reference frames, as summarized in Table \ref{tab:flag}. Despite the initial filtering, numerous bogus sources persisted due to the criteria for each filtering step, which were designed to avoid the loss of real sources. Consequently, point-source-like bogus sources continued to dominate among the filtered sources. Therefore, we took additional steps to clean transient candidates with ML filtering and visual inspection, as described in the following subsections.



\begin{figure*}
    \centering
    \includegraphics[width=1\linewidth]{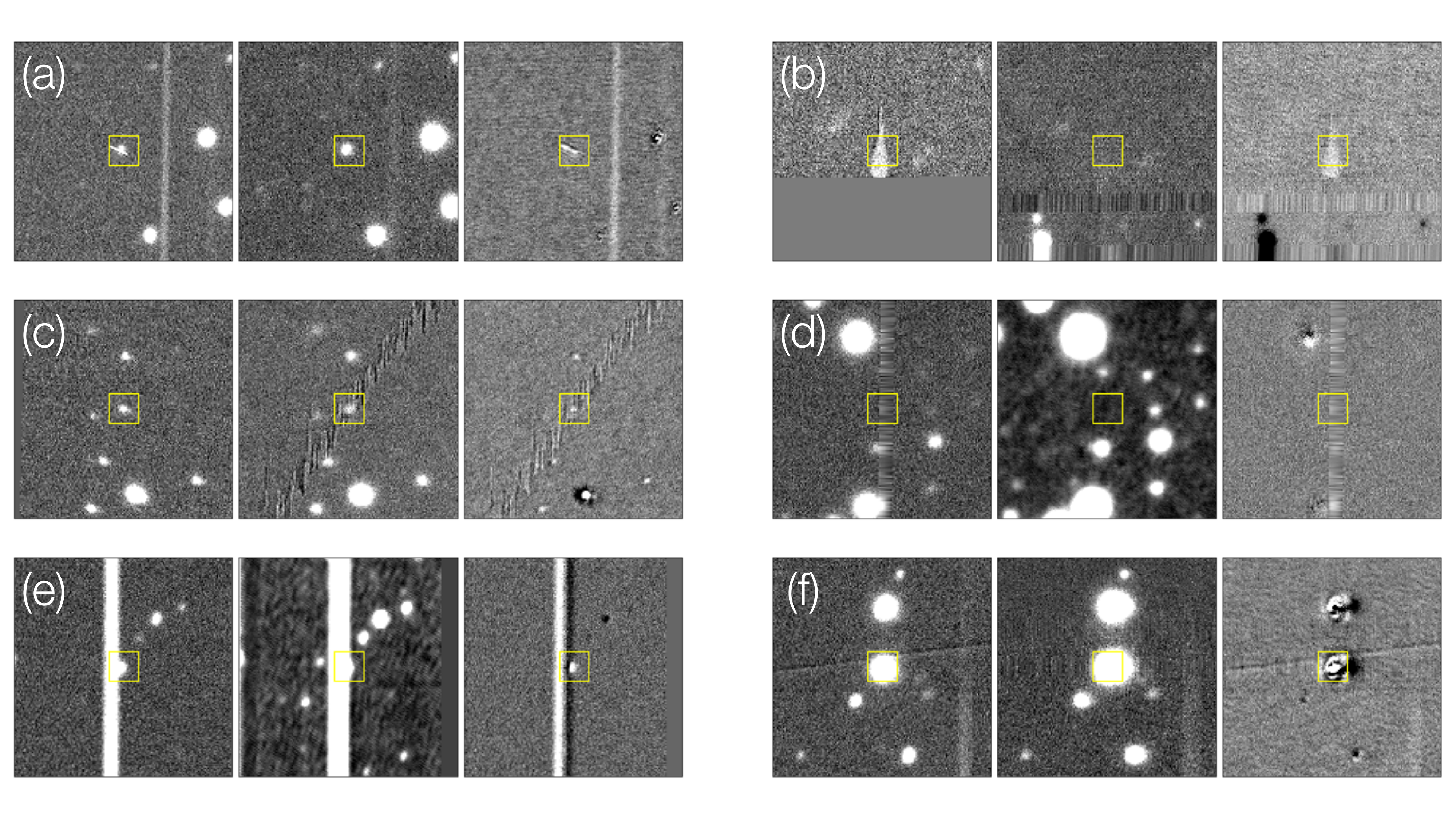}
    \caption{Six representative cases of bogus sources identified through visual inspection. Each set of three images shows the observed, reference, and subtracted frames, with the detected source marked by a yellow box in the center of the image. The cases are described as follows: (a) A cosmic ray in the observed frame, falsely identified as a source. (b) A 'cross-talk' pattern in the observed frame causing a spurious detection. (c) A 'snake-pattern' artifact, stretching from the bottom left to the top right in the reference frame. (d) Column-wise interpolated pixels in the observed frame resulting in a false source. (e) Bleeding from a bright, saturated star visible in both the observed and reference frames. (f) Complex contamination from a 'snake-pattern' in the observed frame, combined with a horizontal interpolation artifact in the reference frame.}
    \label{fig:bad}
\end{figure*}

\begin{figure*}
    \centering
    \includegraphics[width=1.0\textwidth]{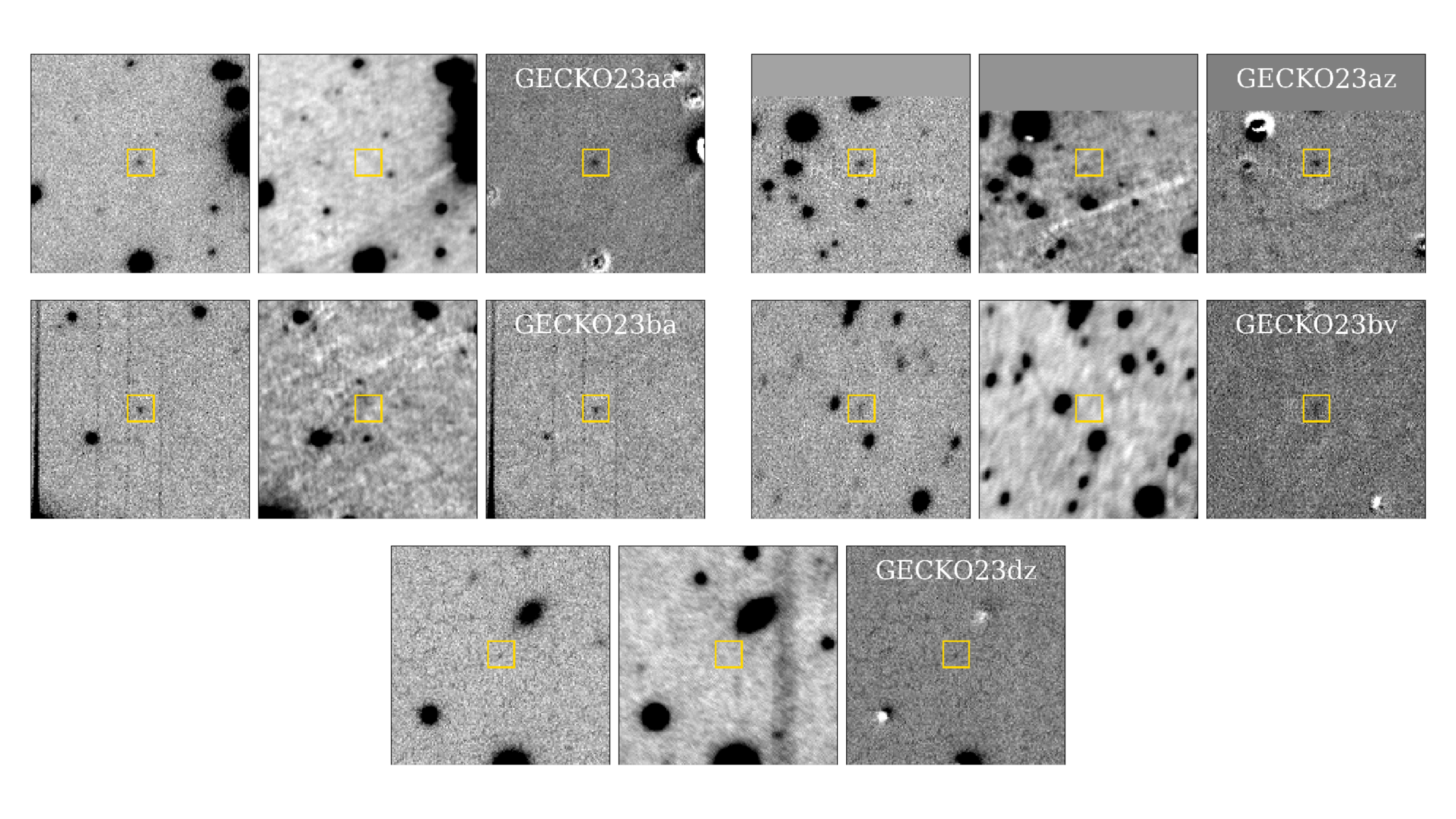}
    \caption{Images of five optical counterpart candidates passing filtering criteria. Left panels: observed images. Middle panels: reference images. Right panels: subtraction images (observed image minus reference image). Yellow boxes: locations of the optical counterpart candidates.}
    \label{fig:candidates}
\end{figure*}

\subsection{Real/Bogus Classifier}\label{ss:rbclassifier}

The structure of our Real/Bogus Classifier is based on the CNN-based \texttt{O'TRAIN} model \citep{2022A&A...664A..81M}. The original \texttt{O'TRAIN} model was designed to use single-channel subtracted images, but we modified it to process three-channel images: observed (sci), reference (ref), and subtracted (sub) images. The input image size is 51 $\times$ 51 pixels, adjusted for feature extraction in our dataset. The classifier was trained using data from the Dark Energy Survey (DES; \citealt{2015AJ....150...82G}), which includes 454,092 real and 444,871 bogus samples. 

The training process is detailed below:

\begin{enumerate}
    \item We split the DES dataset into a training dataset (80\%) and a validation dataset (20\%). The training set included simulated real and bogus samples, while the validation set was used to evaluate the classifier's performance.

    \item To ensure uniform scaling across all input images, each channel underwent L2 normalization. This normalization method adjusts pixel values such that the sum of their squared values equals one, enhancing the convergence of the CNN during training.

    \item The model's input layer was modified to handle three-channel data, integrating information from the sci, ref, and sub images. This modification captures residual patterns in subtraction artifacts, improving the distinction between real and bogus detections.

    \item The classifier's performance was validated using metrics such as accuracy, precision, recall, and F1 score, achieving values exceeding 97\%. However, it should be noted that these metrics were evaluated on DES data and may not directly translate to KMTNet data due to differences in image properties.
\end{enumerate}

This model assigns a probability score ($P_{3ch}$) to each source, indicating the likelihood of being a real transient. For application to KMTNet data, we adopted a loose threshold of $P_{3ch} > 0.5$ to select approximately 57,000 real source candidates from the 476,196 filtered sources. 


However, applying the model to KMTNet data introduces domain-specific challenges. Differences in PSF sizes (DECam: 0.236 arcsec/pixel, KMTNet: 0.4 arcsec/pixel), filter response curves, and unique camera characteristics could lead to performance discrepancies. To address this, our approach involves cross-domain transfer learning to adapt the model for datasets from different facilities. The domain shift is mitigated by retraining the model with a subset of KMTNet data and incorporating adjustments to account for variations in subtraction artifacts.


Future enhancements include expanding the training set with more KMTNet-based data and implementing PSF modeling to generate synthetic sources that closely mimic real transients. These synthetic sources will be injected into the images and used as part of the training set to address the domain differences further. These steps aim to improve the classifier's robustness and ensure more reliable transient detection in future observations.

\subsection{Visual Inspection (KMTNet)}\label{ss:visualinspection}

The filtering process consisted of two steps: an initial inspection using subtracted (1-channel) images, followed by a more detailed inspection using three-channel images (observed, reference, and subtracted). The first step aimed to quickly remove obvious bogus sources, such as cosmic rays and artifacts, facilitating a rapid preliminary selection. However, many sources that appeared genuine during the 1-channel inspection were later identified as bogus due to imperfect subtraction. The second step, utilizing 3-channel images, provided critical context, such as the presence of nearby galaxies in the science image or pre-existing sources in the reference image. This additional information helped to better distinguish real transients from artifacts. The first inspection reduced the candidates to 2,529 sources, and the second inspection ultimately yielded 128 transients, as summarized in Table \ref{tab:candidate}. Representative examples of removed bogus sources are shown in Figure \ref{fig:bad}. These primarily include non-removed cosmic rays, spikes from bright stars poorly corrected through horizontal and vertical interpolation, excessive cross-talk artifacts causing negative sources in the reference image, variable sources, and artificially created sources resulting from significant negative signals in the reference image.




\subsection{Visual Inspection (RASA36)}\label{ss:visualinspection_rasa36}
For RASA36 data, a transient search was performed on 16 tiles. Among these, 13 tiles utilized reference images obtained from RASA36 observations in June 2023, a few weeks after the GW event, while the remaining 3 tiles used $r$-band reference images from the SkyMapper survey \citep{2024PASA...41...61O} for image subtraction because neither RASA36 nor KS4 reference images were available for those areas. Similar to the methods applied for the KMTNet data, first-pass filtering was performed using  \texttt{SExtractor} parameters. Subsequently, filtered transient candidates within 2 arcmins (corresponding to a projected physical distance of approximately 120 kpc at the event's luminosity distance) of the event's host galaxy candidates were visually inspected to determine whether they were real or bogus. Out of 211K detections, 30K (14\%) sources survived during the first-pass filtering, and 1,447 (5\%) transient candidates lie near selected host galaxy candidates. However, no significant final candidates were identified by visual inspection.


\subsection{Optical Counterpart Candidates}\label{ss:opticalcounterparts}

To narrow down plausible optical counterparts among the 128 transients, we set the following criteria:


\begin{enumerate}

    \item The optical counterpart of the GW event should not be known transients: We matched the candidates with objects reported and classified in Transient Name Server (TNS; \citealt{2021AAS...23742305G}) as known transients such as SNe. If we find a matching source with a proper classification, we classify it as so.

    
    \item The KN should be within a projected physical distance of 80 kpc from the nearest host galaxy among matched \texttt{GLADE+} galaxies with localization volume. This involves converting the angular separation between the transient and the host galaxy to a projected physical distance and only selecting those transients that are within 80 kpc. This criterion is based on the maximal projected offsets observed between short GRBs (SGRBs) and their host galaxies, which range from approximately 60 to 75 kpc \citep{2013ApJ...776...18F,2014ARA&A..52...43B,2022ApJ...940...56F}.  To calculate the projected distance, we used the luminosity distances of host galaxies provided from \texttt{GLADE+} catalog. 

    \item The KN, the optical counterpart, is assumed to be dimmer than or similar to the brightness of an AT2017gfo-like KN (Engrave Collaboration; \citealt{2017Natur.551...67P,2017Natur.551...75S}) at the lower limit of the GW luminosity distance in the $r$-band ($d=147.7\:$Mpc). This requirement assumes that the AT2017gfo-like event corresponds to one of the brighter KNe.

    \item The transient should not show past activities (see below).
    

\end{enumerate}





To examine the historical activities of newly identified transients, we employed the ATLAS forced photometry service\footnote{\url{https://fallingstar-data.com/forcedphot/}}, which spans from the year 2000 to May 2024 \citep{2018PASP..130f4505T,2021TNSAN...7....1S}. This service offers forced photometry for specified locations on the historically subtracted ATLAS data. We queried the ATLAS database for these sources and acquired information for 78 of them. However, the forced photometries contain non-astrophysical signals. If evidence of previous activities is found to have occurred more than a few days ago, the transient would likely not be associated with this GW event. Consequently, it is essential to define a parameter to identify significant signals.

First, we counted the number of detections with $\text{SNR} > 5$ ($n_{SNR>5}$), and the defined detection ratio ($R_{SNR>5}$) within a defined time as below: 

$$R_{SNR>5}=n_{SNR>5}/n_{obs}$$

where $n_{obs}$ is the number of objects within a specific time window. To compute these ratios, we divided the time period from $-540$ days to $+120$ days relative to the GW trigger into 30-day intervals and shifted each window by 15 days. For each time window, the number of observations ($n_{obs}$) and the number of detections ($n_{SNR>5}$) were counted. 

\begin{figure*}
    \centering
    \includegraphics[width=0.75\linewidth]{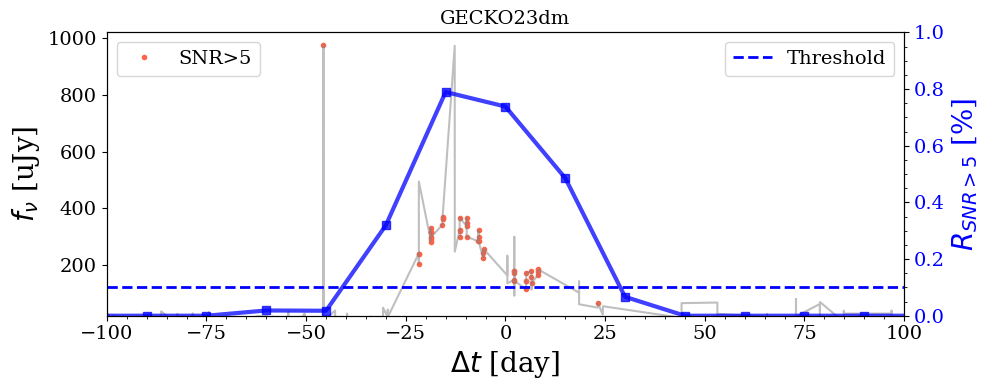}
    \caption{An example of pre-activity for GECKO23dm (AT2023hec), showing significant activity detected using ATLAS forced photometry. The grey line represents the entire photometry dataset, while red circular markers indicate photometry points with an SNR greater than 5. The blue line shows the activity level, defined as the ratio of $\rm SNR > 5$ photometry points to the total number of observations within each time window. The horizontal blue dashed line represents the threshold used to confirm significant activity.}
\label{fig:gecko23dm}
\end{figure*}

An example is shown in Figure \ref{fig:gecko23dm}, illustrating the case of GECKO23dm which exhibited a significant activity a month before the GW trigger. GECKO23dm was already reported as AT2023hec on April 29, 2023, by ATLAS \citep{2023TNSTR.970....1T}.

\begin{figure}
    \centering
    \includegraphics[width=0.5\textwidth]{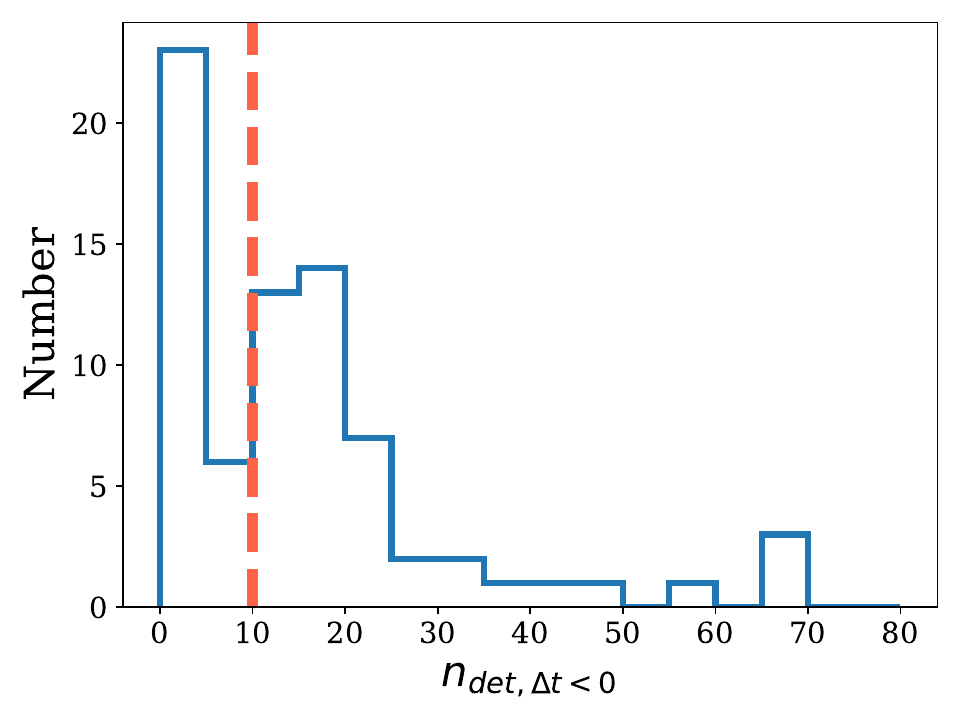}
    \caption{Number distribution of the defined number of activities before GW trigger. The red line shows the threshold to consider the former activity.}
    \label{fig:activity}
\end{figure}

We introduce  $n_{det,\Delta t < 0}$ as the number of time windows with $R_{SNR>5} > 0.01$ (1\%) before the GW trigger. Figure \ref{fig:activity} shows the distribution of $n_{det,\Delta t < 0}$ where it shows a bimodal distribution that can be divided at $n_{det,\Delta t < 0}=10$. Therefore, we consider objects with $n_{det,\Delta t < 0} > 10$ to have had clear pre-activities. As a result, a total of 49 sources exhibited significant activity before the event. 

Among the 128 objects, four bright sources were filtered out by criterion \#1 since they were either reported in TNS well before the GW trigger time or classified by non-KNe by other groups. We found 30 transients satisfying the proximity criterion \#2. We also found 49 sources with pre-event activities (criterion \#4), and 72 sources that are dimmer than the expected bright KN (criterion \#3). 






\begin{figure*}
    \centering
    \includegraphics[width=0.75\linewidth]{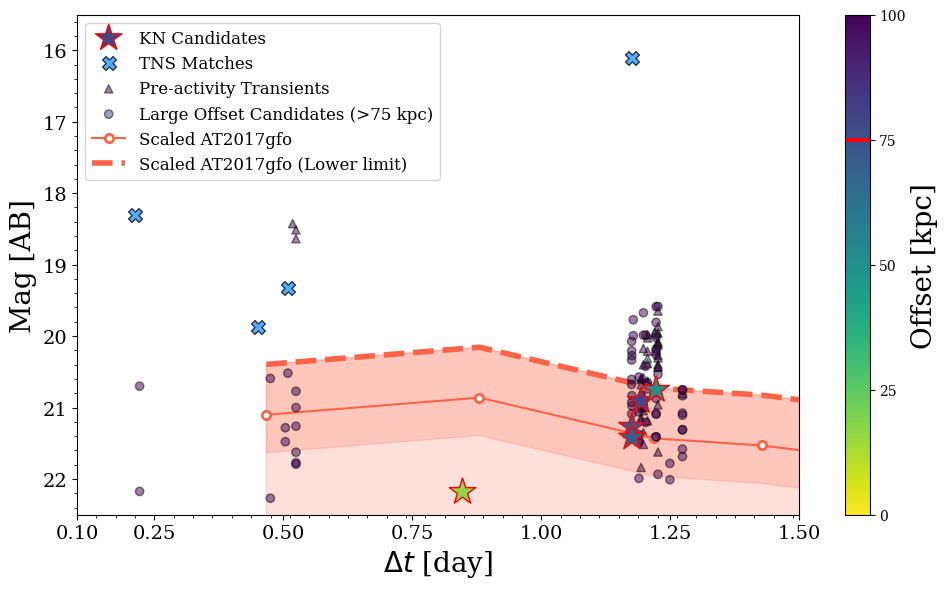}
\caption{Comparison of the photometries of selected transient candidates and the scaled AT2017gfo-like KN light curve at the luminosity distance of S230518h in the $r$-band (red line and shaded area). The KN light curve is derived from the AT2017gfo data release by the Engrave Collaboration \citep{2017Natur.551...67P,2017Natur.551...75S} and scaled to the GW analysis luminosity distance ($\rm204.3 \pm 56.6$ Mpc). The color bar represents the projected offset from each transient to the nearest host galaxy in the matched \texttt{GLADE+} catalog. Star markers indicate KN candidates, which have no pre-activity, no match with TNS, and are within 75 kpc of the projected offset to the nearest host galaxy. Circular markers correspond to distant candidates with offsets greater than 75 kpc, while triangular markers represent transients showing pre-activity before the GW trigger. Cross markers indicate TNS matches, which correspond to transients already reported in the TNS. A red horizontal line in the color bar highlights the 75 kpc threshold for candidate selection.}

\label{fig:transient}
\end{figure*}

\begin{figure}
    \centering
    \includegraphics[width=1.0\linewidth]{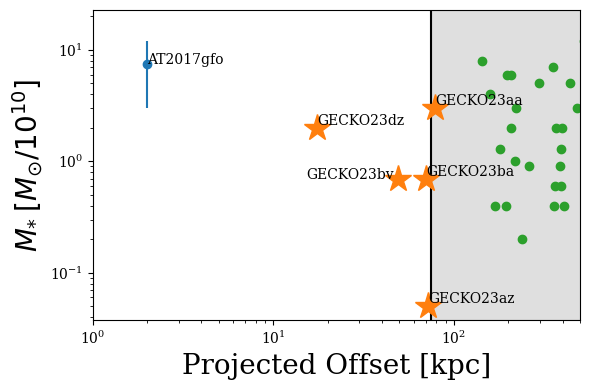}
    \caption{Projected offsets of KN candidates relative to the nearest host galaxy and the stellar masses of host galaxies from the GLADE+ catalog. The projected offset is calculated by converting the projected angular separation between the transient and the host galaxy using the distance from the GLADE+ catalog. The blue marker represents AT2017gfo and its host galaxy, NGC 4993. The orange star markers indicate KN candidates found in the follow-up observation of S230518h. The green circular markers represent transients that are not matched with the TNS and do not show any previous activity, providing additional context for potential candidates. The grey shaded area represents a range larger than 80 kpc, the maximum offset observed in SGRBs.}
    \label{fig:offset_stellar_mass}
\end{figure}


Figure \ref{fig:transient} illustrates the AT2017gfo-like KN light curves scaled to the luminosity distance of S230518h, alongside the single-epoch photometry of the optical counterpart candidates. The figure shows that the photometric points of five KN candidates lie within the allowed range of the scaled light curve or below and satisfy all the criteria. The projected offsets and stellar mass of the host galaxies are shown in Figure \ref{fig:offset_stellar_mass}, where the stellar mass values are taken from the GLADE+ catalog. In terms of the offset, no SGRBs have been found beyond 75 kpc of the projected offset, therefore, it is likely that a KN would lie within this distance from its host galaxy. However, GECKO23aa was included in the analysis despite having a projected offset of 78 kpc, slightly exceeding the 75 kpc threshold, due to its close proximity to this limit. In terms of the host galaxy mass, simulation works show that the host galaxy probability goes with its stellar mass \citep{2018MNRAS.481.5324M,2019ApJ...881L..16A}, and hence, we consider the candidates associated with more massive galaxies to be more likely to be the KN. With these considerations, we find GECKO23dz to be the most plausible KN candidate among the five sources. However, upon a closer inspection of its image, we found that GECKO23dz is likely to be caused by an artifact in the CCD of KMTNet-SSO. While the remaining four candidates meet the preliminary selection criteria, we consider them marginal candidates for the following reasons: (1) as mentioned earlier, they are close to the maximum allowable distance ($\sim$ 75 kpc); (2) A host candidate for GECKO23az has a mass ($\rm < 10^9\:M_{\odot}$) that is too low to be considered a likely host; (3) they were assigned low rankings as they lie at the edges of the GW localization region. Therefore, we conclude that no promising optical counterpart was identified for the GW event.




\section{Discussion}\label{sec:discussion}

\subsection{Constraints on KN Properties}
Although we could not identify a plausible optical counterpart, we can constrain the physical parameters of the expected KN associated with this GW event using theoretical models from the observational upper limits (e.g., \citealt{2020MNRAS.499.3868T} for GW190814 and \citealt{2024ApJ...960..113P} for GW190425). We follow a similar approach to constrain the KN property. We make two key assumptions: first, that the optical counterpart lies within the area covered by KMTNet; second, that we have identified all transients brighter than the 5$\sigma$ depth of the images.  Since different fields were observed at slightly different epochs to various depths (due to sky conditions), plausible models would be those that survive from the upper limit constraint of at least one of the observed fields.

\begin{figure*}
    \centering
    \includegraphics[width=1.0\textwidth]{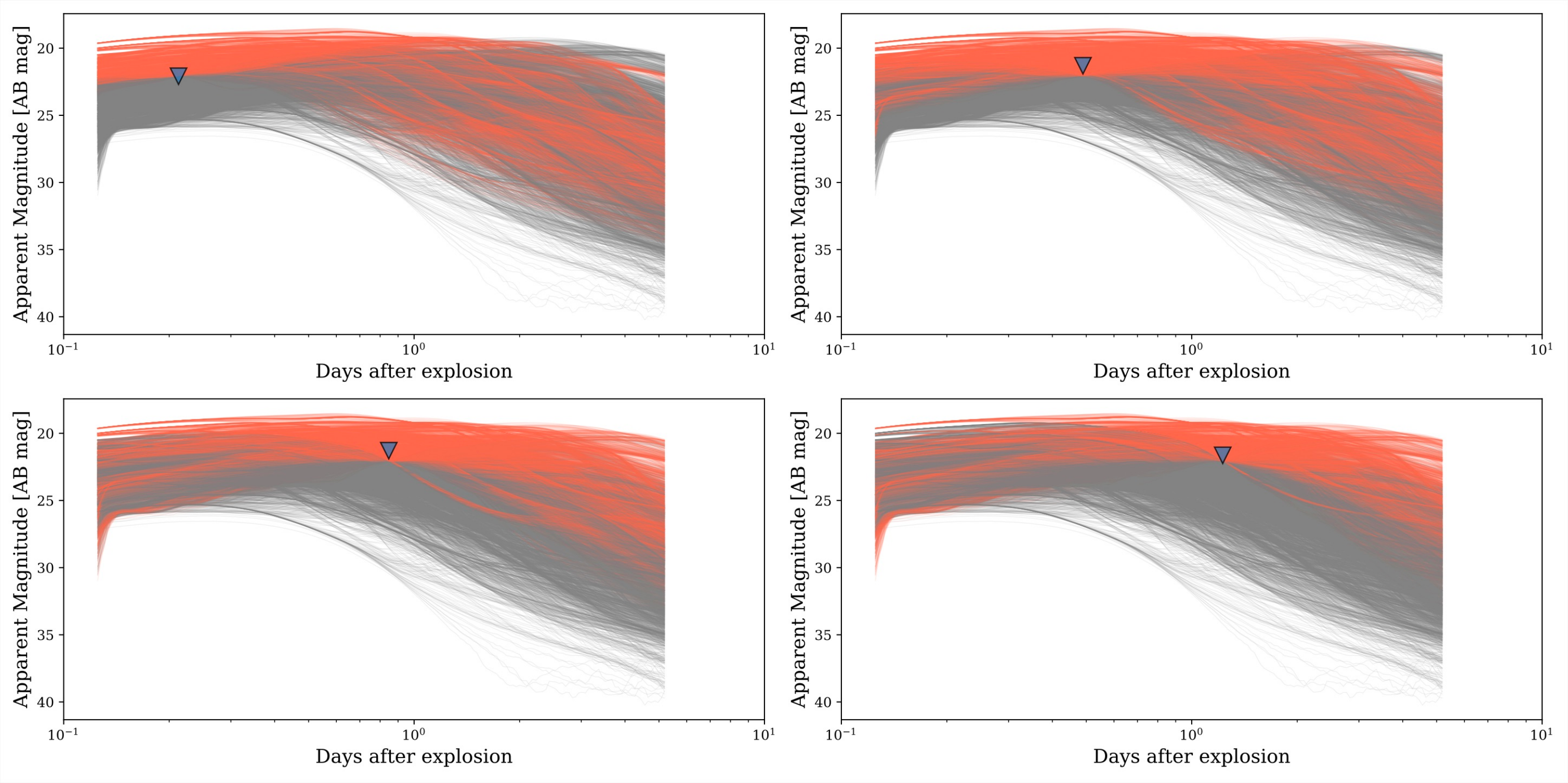}
        \caption{Comparison of 2D KN models with the upper limits of KMTNet follow-up observation of S230518h in $R$-band (blue triangle marker). Each panel shows different regions of the localization area covered at different epochs. Red-colored lines are model LCs ruled out by the upper limits, and the grey-colored lines show models consistent with the upper limits. Among the KN parameter space that produces the red lines in all four panels can be excluded.}
    \label{fig:kn_lc_4epoch}
\end{figure*}

We employed a 2-dimensional grid KN model \citep{2021ApJ...918...10W}, scaled to the median GW luminosity distance of this event ($d=204.3$ Mpc). Free parameters of this model include the viewing angle that varies from on-axis (0 degrees) to edge-on (180 degrees), the dynamical ejecta mass ($\rm m_{dyn}=0.001, 0.003, 0.01, 0.03, 0.1\:M_{\odot}$), the wind ejecta mass ($\rm m_{wind}=0.001, 0.003, 0.01, 0.03, 0.1\:M_{\odot}$), dynamical ejecta velocity ($\rm v_{dyn}=0.05, 0.15, 0.3\:c$), and the wind ejecta velocity ($\rm v_{wind}=0.05, 0.15, 0.3\:c$), and light curves from 9,900 models with different sets of free parameters were produced to match the observational constraints (Figure \ref{fig:kn_lc_4epoch}). Each epoch of KMTNet depth gives independent constraints on the simulated KN light curves as shown in Figure \ref{fig:kn_lc_4epoch}.

\begin{figure}
    \centering
    \includegraphics[width=0.5\textwidth]{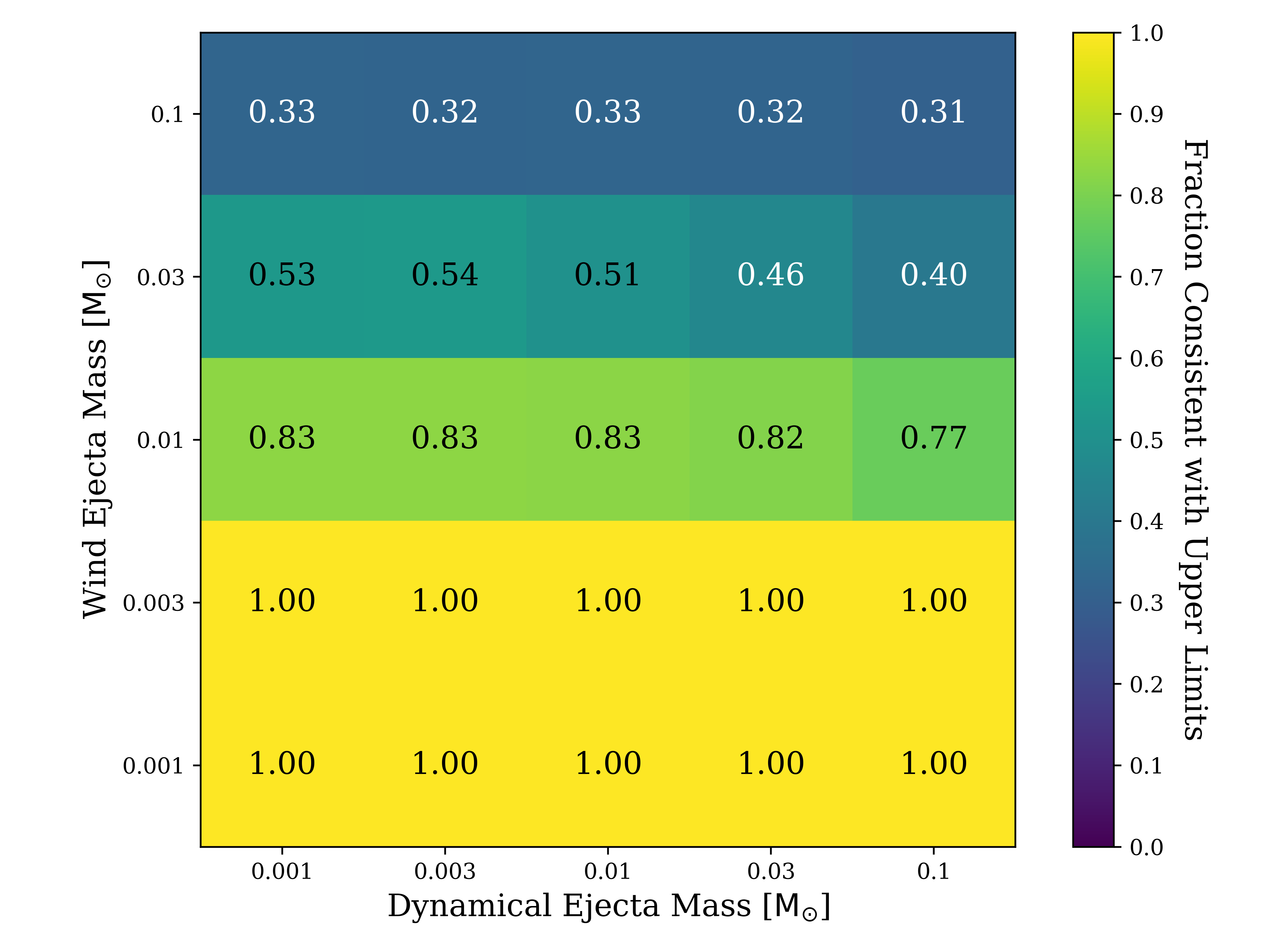}
\caption{Matrix showing the fraction of consistent models with the KMTNet upper limits. Each bin contains 396 models, varying by dynamical ejecta mass (x-axis) and wind ejecta mass (y-axis). The color indicates the fraction: yellower colors signify that most models are consistent with observations (closer to 1.0), while bluer colors indicate that few or none of the models are consistent with observations (closer to 0.0).}
\label{fig:kn_mass_4epoch}
\end{figure}
\begin{figure}
    \centering
    \includegraphics[width=0.5\textwidth]{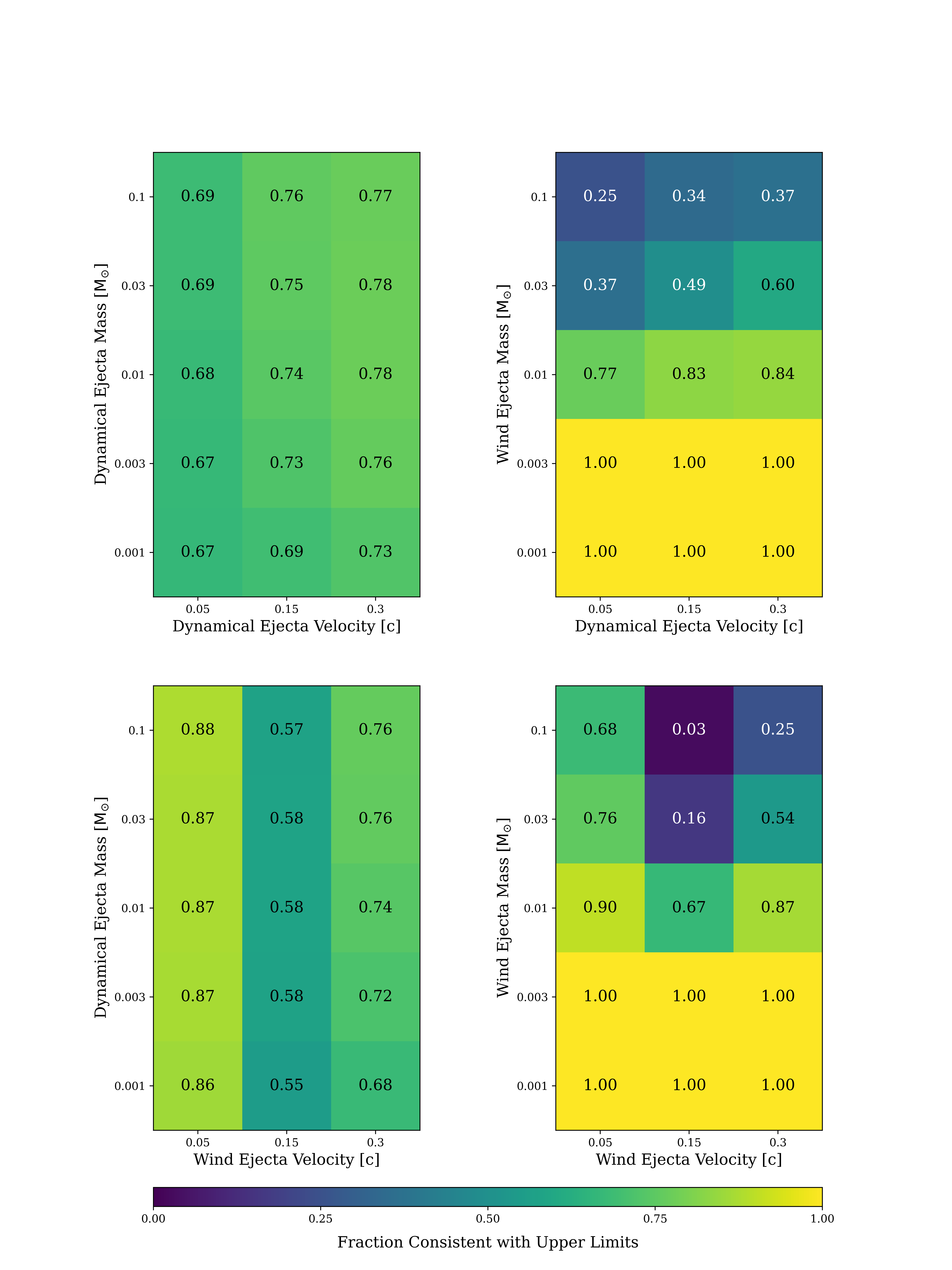}
\caption{Matrix showing the fraction of consistent models with the KMTNet upper limit. Each panel displays combinations of ejecta velocity and mass, with each bin containing 660 models. The color indicates the fraction of consistent models: yellower colors represent a higher consistency with observations, while bluer colors indicate fewer or no consistent models. Bins with fractions less than 0.3, considered less probable, are labeled with white text. (Top left) dynamical ejecta velocity (x-axis) vs. dynamical ejecta mass (y-axis), (Top right) dynamical ejecta velocity (x-axis) vs. wind ejecta mass (y-axis), (Bottom left) wind ejecta velocity (x-axis) vs. dynamical ejecta mass (y-axis), and (Bottom right) wind ejecta velocity (x-axis) vs. wind ejecta mass (y-axis).}
\label{fig:kn_v_m_4epoch}
\end{figure}

As a result, 7,174 models (72.5\%) survived among 9,900. Figures \ref{fig:kn_mass_4epoch} and \ref{fig:kn_v_m_4epoch} show the fraction of the models that survived in matrices of the parameter space. The figures demonstrate that most properties remain weakly constrained by our observations. However, it is plausible that KN models with wind ejecta characterized by very high mass ($\rm m_{wind} > 0.1\:M_{\odot}$) or very high velocity ($\rm v_{wind} > 0.15\:c$) are less likely to be consistent with the observational upper limits, particularly across all viewing angles. Nonetheless, some models in these regimes still exhibit moderate consistency, and thus a definitive exclusion of such models cannot be made.



\subsection{Limitations of the Result}
While the follow-up observations of S230518h provided valuable data, several limitations and challenges were encountered that impacted the overall results. 

\begin{figure}
    \centering
    \includegraphics[width=1\linewidth]{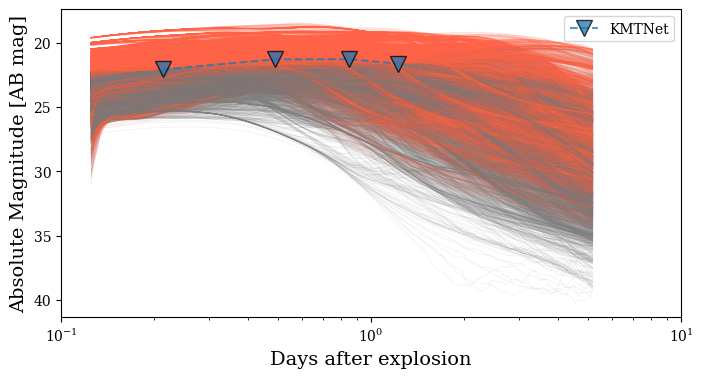}
        \caption{Comparison of 2D KN models with the upper limit of KMTNet follow-up observation of S230518h in $R$-band (blue triangle marker), assuming that the same area was covered in each epoch. Each panel shows different epochs of observation. Red-colored lines are consistent model LCs with the upper limit, and grey-colored lines are consistent model LCs.}
    \label{fig:lc_whatif}
\end{figure}
\begin{figure}
    \centering
    \includegraphics[width=1\linewidth]{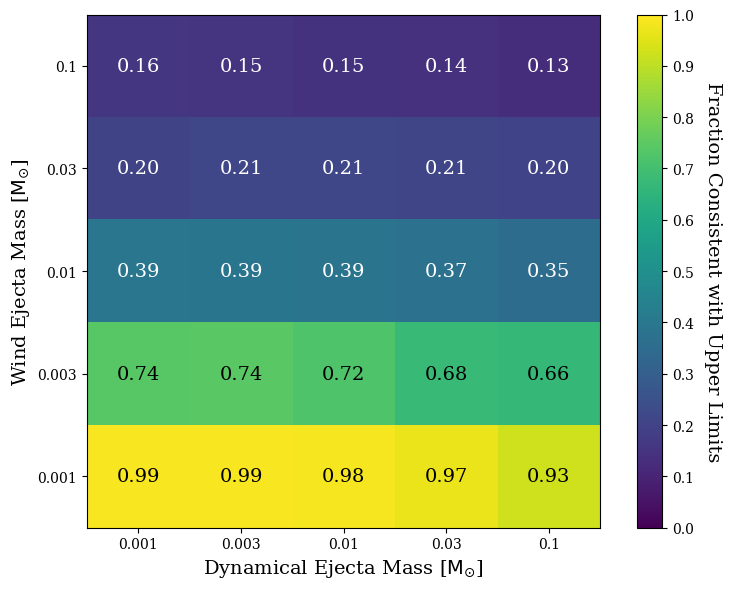}
\caption{Matrix showing the fraction of consistent models with the KMTNet upper limits, assuming that the same area was covered in each epoch. Each bin contains 396 models, varying by dynamical ejecta mass (x-axis) and wind ejecta mass (y-axis). The color indicates the fraction: yellower colors signify that most models are consistent with observations (closer to 1.0), while bluer colors indicate that few or none of the models are consistent with observations (closer to 0.0). Bins with a fraction less than 0.3 are considered less probable models and are labeled with white-colored text.}
    \label{fig:ejecta_mass_whatif}
\end{figure}
\begin{figure}
    \centering
    \includegraphics[width=1\linewidth]{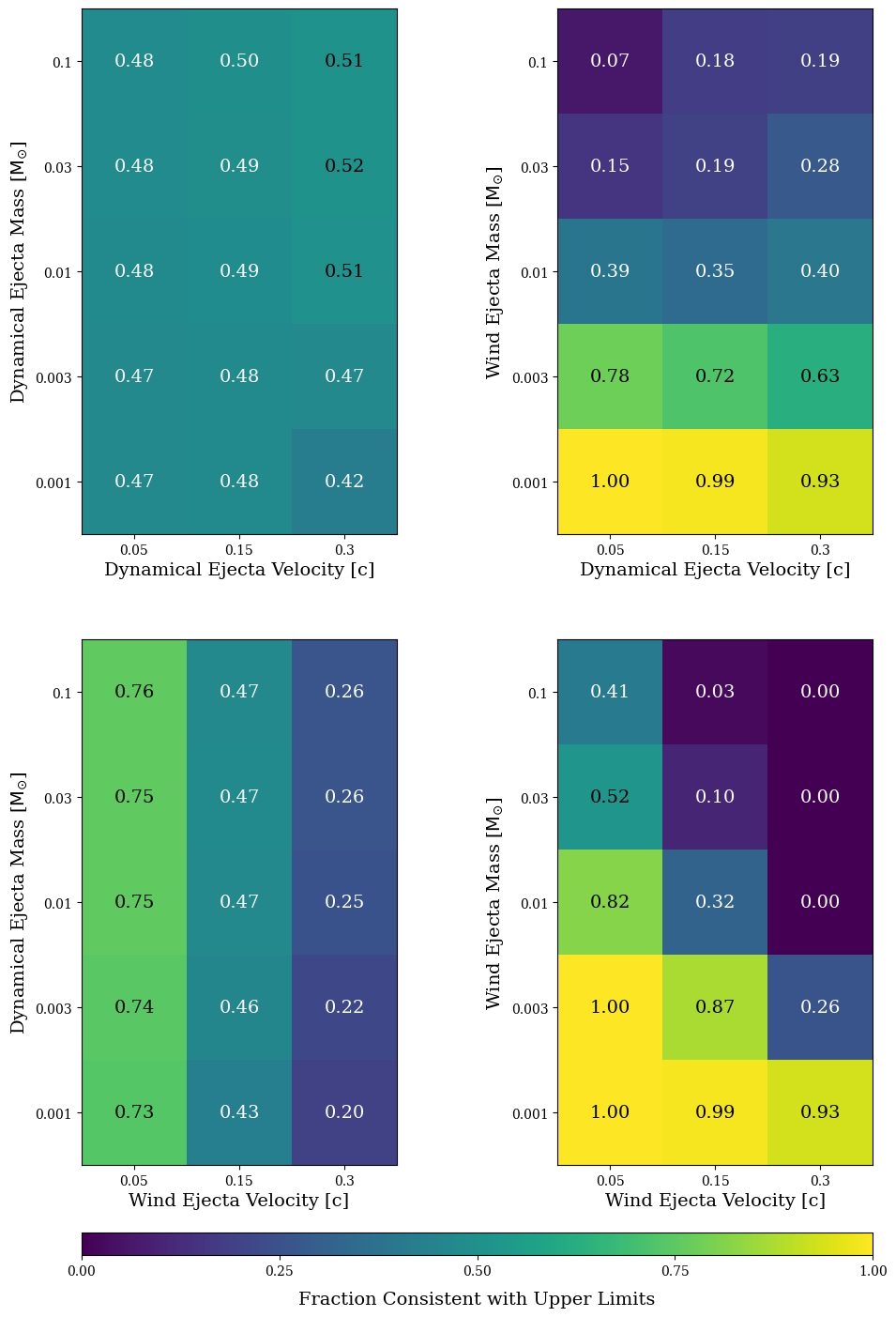}
\caption{Matrix showing the fraction of consistent models with the KMTNet upper limit, assuming that the same area was covered in each epoch. Each panel displays combinations of ejecta velocity and mass, with each bin containing 660 models. The color indicates the fraction of consistent models: yellower colors represent a higher consistency with observations, while bluer colors indicate fewer or no consistent models. Bins with fractions less than 0.3, considered less probable, are labeled with white text. (Top left) dynamical ejecta velocity (x-axis) vs. dynamical ejecta mass (y-axis), (Top right) dynamical ejecta velocity (x-axis) vs. wind ejecta mass (y-axis), (Bottom left) wind ejecta velocity (x-axis) vs. dynamical ejecta mass (y-axis), and (Bottom right) wind ejecta velocity (x-axis) vs. wind ejecta mass (y-axis).}    \label{fig:ejecta_mass_vel_whatif}
\end{figure}

1. The absence of multi-epoch and multi-filter observations limited our ability to classify transient candidates. When using only photometric data, tracing the evolution of light curves and color, which are critical for distinguishing between different types of transients, and constraining KN properties. Figure \ref{fig:lc_whatif} shows the expected constraints if we had observed each field at four different epochs. Compared with the single epoch result (Figure \ref{fig:kn_lc_4epoch}), 4,780 (48.3\%) models survived and the constraint on the wind ejecta is much tighter ($m_{wind}<0.03\:M_{\odot}$ and $v_{wind} < 0.15\:c$) as shown in Figures \ref{fig:ejecta_mass_whatif} and \ref{fig:ejecta_mass_vel_whatif}. This simulation suggests that multiple visits would have significantly improved our constraints on the KN properties.


2. The observational window was constrained by the position of the Sun, which created an unobservable gap in the northern hemisphere. This reduced the coverage and the usefulness of the GECKO data to constrain KN properties since the optical counterpart could have existed in the northern unobservable gap.

3. In the $K$-band, the \texttt{GLADE+} catalog contains 100 and 85\% of the brightest galaxies that produce 90\% of the total luminosity at $<$ 130 Mpc and $\sim$200 Mpc \citep{2022MNRAS.514.1403D}. Since BNS merger rates are expected to strongly correlate with the host galaxy mass (or $K$-band luminosity as a proxy), we expect only a modest incompleteness in the KN host candidates list. However, if the BNS merger rates weakly correlate with the galaxy mass, the incompleteness in the \texttt{GLADE+} catalog can be considered significant (50\% at 200 Mpc in $B$-band), in which case our constraints on KN properties should be examined further with a more comprehensive catalog.

\subsection{Improving GECKO Follow-up Strategy}

Based on our experience with S230518h, we analyze GECKO's capabilities and identify areas for improvement in future follow-up campaigns. As evidenced by the follow-up observations of GW190425 \citep{2024ApJ...960..113P}, tiling observations with KMTNet, a wide-FoV telescope, outperform galaxy-targeted observations in terms of coverage and efficiency for events with large localization errors. We investigate the efficiency of KMTNet in following up S230518h-like events with a luminosity distance of 204 Mpc in terms of multi-epoch observation. Several assumptions are made:

\begin{enumerate}
    \item The entire CR90 region is observable.
    \item The overhead time between pointings is 12 minutes, with an exposure time of $\rm t_{exp} = 120s\times4$.
    \item All KMTNet facilities can be used for the observation.
    \item The optical counterpart is similar to an AT2017gfo-like KN.
\end{enumerate}



The sampling rate of the KN light curve depends on the size of CR90 ($\Omega_{\text{CR90}}$), observation time per field ($t_{\text{obs}}$), and the FOV of KMTNet ($\mathrm{FOV} \sim 2 \times 2\:\mathrm{deg}^2$).

The observation time per field can be expressed as:
\[
t_{\text{obs}} = t_{\text{exp}} \times N_{\text{frames}} + t_{\text{readout}} \times N_{\text{frames}} + t_{\text{overhead}},
\]
where the on-source exposure time per frame is $t_{\text{exp}} = 2\:\mathrm{min}$, and the number of frames per field is $N_{\text{frames}} = 4$. The readout time for each frame is $t_{\text{readout}} = 30\:\mathrm{s}$, and the additional overhead time for telescope pointing, focusing, and filter changes is $t_{\text{overhead}} = 2\:\mathrm{min}$. Thus, the total observation time per field is $t_{\text{obs}} = 12\:\mathrm{min}$.

The total observation time, $T_{\text{obs}}$ can be calculated as follows, including an efficiency factor $\eta$ to account for overlap inefficiencies:


\[
T_{\text{obs}} = \frac{t_{\text{obs}} \times \Omega_{\text{CR90}}}{\text{FOV}} \times \eta
\]


The efficiency factor ($\eta \sim 1.2$) reflects the mismatch between the rectangular KMTNet tiles and the typically elliptical shape of the CR90 region. KMTNet uses a predefined tiling scheme that is not optimized for the random shapes of GW localization areas. For example, a simple calculation based on the size of KMTNet’s FOV and the area of the S230518h CR90 region suggests that approximately 115 tiles would be required. However, due to the predefined tile arrangement and necessary overlap, 139 tiles are actually needed to effectively cover the CR90 region. For an area of 460 deg$^2$, the total observation time to cover the whole area with 3 KMTNet telescopes is about 24 hours, which can be covered by KMTNet observations at all three sites if the target is in the visible window for the whole night.

\begin{figure}
    \centering
    \includegraphics[width=1.0\linewidth]{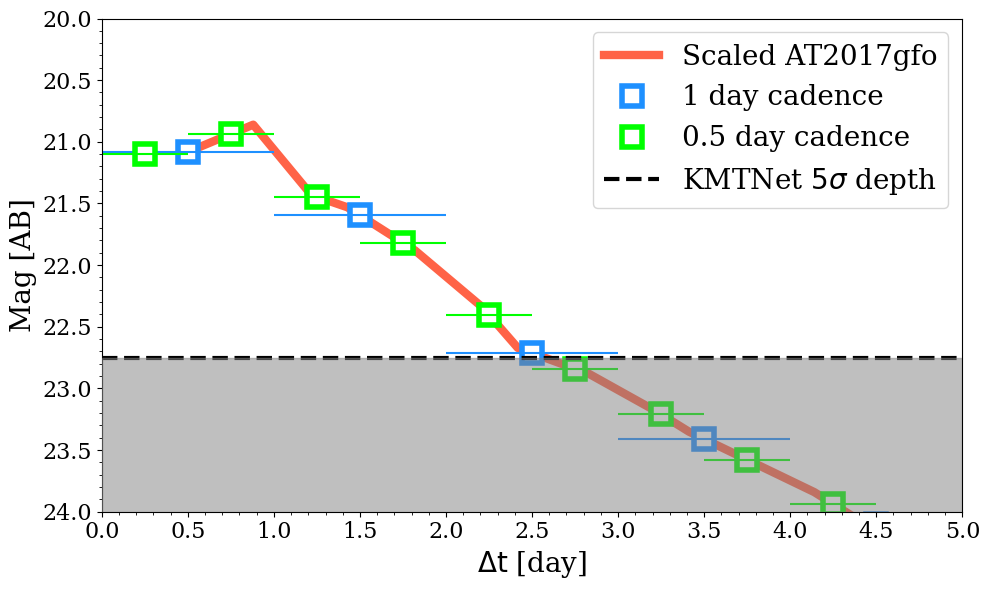}
    \caption{Scaled light curve of an AT2017gfo-like kilonova at 204 Mpc, showing sampling rates with 1-day (blue) and 0.5-day (green) cadences. The KMTNet 5$\sigma$ depth is indicated as a dashed line.}
    \label{fig:whatif_sampling}
\end{figure}

The median 5$\sigma$ depth of KMTNet with 8 min on-source exposure time, in the $R$-band at 12 min integration time is approximately 22.75 mag \citep{2021ApJ...916...47K,2024ApJ...960..113P}. Figure \ref{fig:whatif_sampling} illustrates the light curve of a scaled AT2017gfo-like KN with respect to the KMTNet depth. Assuming a GW event similar to S230518h ($\Omega_{\text{CR90}} \leq 460 \:\text{deg}^2$), a 1-day cadence can cover the entire CR90 area, enabling 2–3 multi-epoch observations. For smaller regions (e.g., $\Omega_{\text{CR90}} \leq 230 \:\text{deg}^2$), 5 multi-epoch observations are feasible, or alternatively, multi-band (e.g., $B$ and $R$) observations can be achieved with a 1-day cadence.

The ML application effectively reduced the number of transient candidates detected in KMTNet data. Approximately 88\% of the transient candidates were rejected using a conservative threshold ($\rm P_{3ch} > 0.5$). However, even after the ML filtering, the average number of transient candidates per image remains around 1,000 before visual inspection. Additionally, while the RB classifier itself takes only a few minutes to execute, the overall processing time per image was dominated by other steps, including image stacking, photometry, subtraction, and generating stamp images from the subtracted image, resulting in a total processing time of approximately 60 to 90 minutes per a single image.

To further improve the efficiency of transient searches, we plan to optimize both the data processing time and the model tuning for the KMTNet data by preparing GPU-based computational hardware and software, upgrading the RB-classifier, and simplifying the classifier workflow.

\section{Summary}\label{sec:summaryt}

We conducted optical follow-up observations of the NSBH GW event, S230518h, using the GECKO telescope network. This event had a localization area (CR90) of 460.1 deg$^2$ and a distance estimate of $204.3 \pm 57$ Mpc, providing an excellent opportunity to test the effectiveness of the GECKO facilities and our data reduction and transient search pipeline. Observations were performed in the $R$-band with KMTNet, the $r$-band with RASA36, and the $r$-band with LSGT covering nearly all of the CR90 localization area observable from the southern hemisphere. The total coverage amounted to 284 deg$^2$, or 61.7\% of the CR90 area, using 71 tiles.

During the transient search process, we initially detected approximately 13 million candidates through image subtraction. This large number was systematically reduced by applying a series of filtering criteria, leaving 128 transient candidates. These candidates were further evaluated using a CNN-based real/bogus classifier, \texttt{O'TRAIN} \citep{2022A&A...664A..81M}, followed by visual inspection. Five transients satisfied our criteria for identifying an optical counterpart, but one was determined to be an artifact, resulting in four remaining candidates. Due to the limitations of single-epoch and single-band data, no further confirmation or detailed analysis could be conducted, and these candidates were retained for future study.

The follow-up observations demonstrated that GECKO can begin observations within a few hours of receiving an alert, achieving a coverage rate of 20 deg$^2$ per hour at a limiting depth of $R \sim 22$ mag. However, our findings suggest that waiting for the \texttt{INITIAL} Bayestar alert can significantly reduce the search area (by nearly a factor of two), improving overall efficiency.

Although limited to single-epoch and single-band data, our observations place marginal constraints on the KN properties, with the wind ejecta mass estimated to be less than 0.1 $M_{\odot}$ and the wind ejecta velocity $< 0.3\:c$ if the GW event occurred in the observed region. These constraints could be improved approximately twofold with multi-epoch observations spanning three to four epochs.

The transient search process has significantly improved since O3, with data reduction and candidate identification times reduced from weeks to days during O4a. Current efforts focus on further improving the pipeline speed by training the CNN-based classifier with additional test data and incorporating GPU-based data processing.

Finally, following this event, we integrated the 7-dimensional telescope (7DT) into the GECKO network. This new system comprises 20 wide-field 0.5-m telescopes and offers three observation modes: spectral mapping (400 to 900 nm) over a FoV of 1.2 deg$^2$ at a spectral resolution of $\sim 40$, deep imaging of the same field of view with an equivalent 2.3-m diameter aperture, and 25 deg$^2$ wide-field imaging with a single filter \citep{2021cosp...43E1537I,2024ApJ...960..113P,2024SPIE13094E..0XK}. With the addition of 7DT and further pipeline enhancements, we expect to achieve significantly improved efficiency in GW optical counterpart searches during the remaining O4 period and beyond.

\begin{longtable*}{ccccccccccc}
    \caption{Transients found by GECKO} \\
    \toprule
    GECKO ID & R.A. & Dec. & Mag. & Mag. Error & DATE-OBS & $\Delta t$ & confidence & $D_{L}$ & $\theta_{galaxy}$ & rank \\
             & deg  & deg  & ABmag & ABmag & UTC & s & \% & Mpc & arcsecond & \\
    \hline
    \endfirsthead

    \caption[]{Transient candidates observed by GECKO (continued)} \\
    \toprule
    GECKO ID & R.A. & Dec. & Mag. & Mag. Error & DATE-OBS & $\Delta t$ & confidence & $D_{L}$ & $\theta_{galaxy}$ & rank \\
             & deg  & deg  & ABmag & ABmag & UTC & s & \% & Mpc & arcsecond & \\
    \hline
    \endhead

    \hline
    \multicolumn{11}{p{\textwidth}}{\footnotesize \textbf{Note:} Objects in boldface are transients with potential GW host galaxies within a projected distance of 80 kpc. Among these, GECKO23dz is found spurious due to special artifacts in the KMTNet-SSO image.} \\
    \endfoot

    \hline
    \endlastfoot

    GECKO23a & 105.153 & -28.651 & 20.63 & 0.12 & 2023-05-19T17:55:17 & 1.206 & 0.99 & 241.1 & 250.2 & 7985 \\
    GECKO23b & 104.501 & -28.503 & 20.38 & 0.10 & 2023-05-19T17:55:17 & 1.206 & 0.99 & 213.8 & 379.3 & 8249 \\
    GECKO23c & 104.474 & -28.474 & 21.19 & 0.19 & 2023-05-19T17:55:17 & 1.206 & 0.95 & 189.9 & 392.2 & 5767 \\
    GECKO23d & 104.429 & -28.219 & 20.32 & 0.07 & 2023-05-19T17:55:17 & 1.206 & 0.99 & 203.5 & 63.0 & 6032 \\
    GECKO23e & 104.427 & -28.180 & 20.03 & 0.09 & 2023-05-19T17:55:17 & 1.206 & 0.99 & 203.5 & 111.2 & 6032 \\
    GECKO23f & 104.500 & -28.203 & 19.98 & 0.08 & 2023-05-19T17:55:17 & 1.206 & 0.9 & 248.0 & 98.8 & 932 \\
    GECKO23g & 104.488 & -28.108 & 20.21 & 0.07 & 2023-05-19T17:55:17 & 1.206 & 0.99 & 298.8 & 80.1 & 6939 \\
    GECKO23h & 104.413 & -28.055 & 20.65 & 0.12 & 2023-05-19T17:55:17 & 1.206 & 0.99 & 240.5 & 155.4 & 6651 \\
    GECKO23i & 104.523 & -28.022 & 20.66 & 0.11 & 2023-05-19T17:55:17 & 1.206 & 0.99 & 186.1 & 144.1 & 7140 \\
    GECKO23j & 104.440 & -27.919 & 20.74 & 0.11 & 2023-05-19T17:55:17 & 1.206 & 0.95 & 270.3 & 179.9 & 5217 \\
    GECKO23k & 106.685 & -28.655 & 19.59 & 0.05 & 2023-05-19T18:24:24 & 1.226 & 1.0 & 266.6 & 2403.6 & 10832 \\
    GECKO23l & 106.610 & -28.653 & 19.65 & 0.08 & 2023-05-19T18:24:24 & 1.226 & 1.0 & 266.6 & 2164.3 & 10832 \\
    GECKO23m & 106.706 & -28.575 & 20.96 & 0.10 & 2023-05-19T18:24:24 & 1.226 & 1.0 & 266.6 & 2475.8 & 10832 \\
    GECKO23n & 106.682 & -28.443 & 21.62 & 0.13 & 2023-05-19T18:24:24 & 1.226 & 1.0 & 266.6 & 2487.0 & 10832 \\
    GECKO23o & 106.681 & -28.137 & 20.27 & 0.06 & 2023-05-19T18:24:24 & 1.226 & 1.0 & 266.6 & 2983.1 & 10832 \\
    GECKO23p & 106.615 & -28.074 & 20.40 & 0.07 & 2023-05-19T18:24:24 & 1.226 & 1.0 & 266.6 & 2967.3 & 10832 \\
    GECKO23q & 106.707 & -28.011 & 20.44 & 0.11 & 2023-05-19T18:24:24 & 1.226 & 1.0 & 266.6 & 3333.7 & 10832 \\
    GECKO23r & 106.587 & -27.941 & 19.87 & 0.07 & 2023-05-19T18:24:24 & 1.226 & 1.0 & 266.6 & 3250.9 & 10832 \\
    GECKO23s & 106.660 & -27.941 & 20.10 & 0.08 & 2023-05-19T18:24:24 & 1.226 & 1.0 & 266.6 & 3405.9 & 10832 \\
    GECKO23t & 106.601 & -27.923 & 19.96 & 0.05 & 2023-05-19T18:24:24 & 1.226 & 1.0 & 266.6 & 3329.7 & 10832 \\
    GECKO23u & 106.585 & -27.811 & 20.08 & 0.10 & 2023-05-19T18:24:24 & 1.226 & 0.99 & 267.1 & 3309.0 & 7406 \\
    GECKO23v & 106.632 & -27.782 & 20.43 & 0.10 & 2023-05-19T18:24:24 & 1.226 & 0.99 & 267.1 & 3449.2 & 7406 \\
    GECKO23w & 106.682 & -27.746 & 20.30 & 0.09 & 2023-05-19T18:24:24 & 1.226 & 0.99 & 267.1 & 3601.3 & 7406 \\
    GECKO23x & 106.613 & -27.694 & 20.12 & 0.09 & 2023-05-19T18:24:24 & 1.226 & 0.99 & 267.1 & 3383.8 & 7406 \\
    GECKO23y & 105.131 & -29.649 & 21.93 & 0.15 & 2023-05-19T18:24:24 & 1.226 & 0.9 & 273.6 & 473.1 & 3785 \\
    GECKO23z & 105.325 & -27.685 & 20.54 & 0.10 & 2023-05-19T18:24:24 & 1.226 & 1.0 & 197.3 & 149.7 & 11866 \\
    \textbf{GECKO23aa} & 102.722 & -26.616 & 20.91 & 0.10 & 2023-05-19T17:37:22 & 1.193 & 1.0 & 72.4 & 224.8 & 13322 \\
    GECKO23ab & 102.685 & -26.489 & 20.97 & 0.09 & 2023-05-19T17:37:22 & 1.193 & 0.95 & 227.3 & 187.5 & 4240 \\
    GECKO23ac & 102.642 & -26.373 & 20.59 & 0.10 & 2023-05-19T17:37:22 & 1.193 & 0.9 & 322.0 & 60.5 & 914 \\
    GECKO23ad & 102.663 & -26.350 & 21.01 & 0.10 & 2023-05-19T17:37:22 & 1.193 & 0.9 & 322.0 & 50.8 & 914 \\
    GECKO23ae & 102.764 & -25.956 & 20.61 & 0.10 & 2023-05-19T17:37:22 & 1.193 & 0.9 & 301.5 & 141.0 & 2767 \\
    GECKO23af & 102.702 & -25.734 & 21.15 & 0.13 & 2023-05-19T17:37:22 & 1.193 & 0.99 & 312.7 & 374.7 & 8440 \\
    GECKO23ag & 102.691 & -25.688 & 20.79 & 0.09 & 2023-05-19T17:37:22 & 1.193 & 0.95 & 208.8 & 304.4 & 4300 \\
    GECKO23ah & 102.831 & -25.649 & 21.84 & 0.11 & 2023-05-19T17:37:22 & 1.193 & 0.9 & 267.5 & 227.9 & 3066 \\
    GECKO23ai & 102.719 & -25.681 & 21.18 & 0.09 & 2023-05-19T17:37:22 & 1.193 & 0.95 & 167.5 & 345.4 & 5304 \\
    GECKO23aj & 102.646 & -25.685 & 20.68 & 0.10 & 2023-05-19T17:37:22 & 1.193 & 0.95 & 208.8 & 228.3 & 4300 \\
    GECKO23ak & 105.397 & -26.627 & 21.52 & 0.11 & 2023-05-19T17:43:46 & 1.198 & 0.99 & 174.6 & 1467.6 & 8422 \\
    GECKO23al & 104.823 & -26.729 & 20.93 & 0.12 & 2023-05-19T17:43:46 & 1.198 & 1.0 & 237.8 & 497.0 & 10270 \\
    GECKO23am & 104.907 & -26.718 & 21.35 & 0.11 & 2023-05-19T17:43:46 & 1.198 & 0.99 & 162.9 & 679.4 & 8882 \\
    GECKO23an & 104.746 & -26.715 & 20.40 & 0.05 & 2023-05-19T17:43:46 & 1.198 & 1.0 & 237.8 & 509.4 & 10270 \\
    GECKO23ao & 104.788 & -26.704 & 20.91 & 0.08 & 2023-05-19T17:43:46 & 1.198 & 1.0 & 237.8 & 554.2 & 10270 \\
    GECKO23ap & 104.765 & -26.671 & 20.58 & 0.09 & 2023-05-19T17:43:46 & 1.198 & 0.9 & 156.2 & 580.8 & 3914 \\
    GECKO23aq & 104.846 & -26.554 & 21.04 & 0.12 & 2023-05-19T17:43:46 & 1.198 & 0.9 & 156.2 & 940.1 & 3914 \\
    GECKO23ar & 104.703 & -26.372 & 20.99 & 0.13 & 2023-05-19T17:43:46 & 1.198 & 1.0 & 177.9 & 937.2 & 11540 \\
    GECKO23as & 104.801 & -26.344 & 20.17 & 0.07 & 2023-05-19T17:43:46 & 1.198 & 1.0 & 177.9 & 983.8 & 11540 \\
    GECKO23at & 104.748 & -25.803 & 20.91 & 0.12 & 2023-05-19T17:43:46 & 1.198 & 1.0 & 148.5 & 390.1 & 12646 \\
    GECKO23au & 104.821 & -25.716 & 20.83 & 0.11 & 2023-05-19T17:43:46 & 1.198 & 1.0 & 148.5 & 767.9 & 12646 \\
    GECKO23av & 104.760 & -25.689 & 19.67 & 0.06 & 2023-05-19T17:43:46 & 1.198 & 1.0 & 151.6 & 712.9 & 11492 \\
    GECKO23aw & 104.782 & -25.627 & 19.99 & 0.06 & 2023-05-19T17:43:46 & 1.198 & 1.0 & 151.6 & 873.7 & 11492 \\
    GECKO23ax & 104.807 & -25.683 & 20.91 & 0.09 & 2023-05-19T17:43:46 & 1.198 & 1.0 & 148.5 & 823.7 & 12646 \\
    GECKO23ay & 102.918 & -24.629 & 20.98 & 0.12 & 2023-05-19T17:11:47 & 1.175 & 0.99 & 279.4 & 357.3 & 9132 \\
    \textbf{GECKO23az} & 102.893 & -24.287 & 21.27 & 0.11 & 2023-05-19T17:11:47 & 1.175 & 0.99 & 209.3 & 71.0 & 7879 \\
    \textbf{GECKO23ba} & 102.904 & -24.324 & 21.42 & 0.14 & 2023-05-19T17:11:47 & 1.175 & 0.99 & 209.3 & 69.4 & 7879 \\
    GECKO23bb & 102.967 & -24.025 & 20.27 & 0.08 & 2023-05-19T17:11:47 & 1.175 & 0.99 & 185.1 & 238.7 & 6979 \\
    GECKO23bc & 103.026 & -23.757 & 20.83 & 0.09 & 2023-05-19T17:11:47 & 1.175 & 1.0 & 138.7 & 329.4 & 12527 \\
    GECKO23bd & 105.068 & -24.795 & 19.59 & 0.07 & 2023-05-19T18:19:11 & 1.222 & 0.95 & 160.3 & 2597.6 & 5611 \\
    GECKO23be & 105.009 & -24.741 & 20.75 & 0.11 & 2023-05-19T18:19:11 & 1.222 & 0.95 & 160.3 & 2572.7 & 5611 \\
    GECKO23bf & 105.034 & -24.740 & 20.77 & 0.13 & 2023-05-19T18:19:11 & 1.222 & 0.95 & 160.3 & 2636.1 & 5611 \\
    GECKO23bg & 105.044 & -24.662 & 20.18 & 0.10 & 2023-05-19T18:19:11 & 1.222 & 0.95 & 160.3 & 2853.9 & 5611 \\
    GECKO23bh & 104.685 & -24.572 & 21.41 & 0.18 & 2023-05-19T18:19:11 & 1.222 & 0.99 & 225.4 & 2347.0 & 8879 \\
    GECKO23bi & 105.083 & -24.530 & 20.77 & 0.10 & 2023-05-19T18:19:11 & 1.222 & 0.95 & 160.3 & 3288.4 & 5611 \\
    GECKO23bj & 104.970 & -24.399 & 19.99 & 0.08 & 2023-05-19T18:19:11 & 1.222 & 0.99 & 225.4 & 3447.3 & 8879 \\
    GECKO23bk & 105.041 & -24.318 & 19.81 & 0.06 & 2023-05-19T18:19:11 & 1.222 & 0.95 & 227.7 & 3656.3 & 4463 \\
    GECKO23bl & 104.983 & -24.333 & 20.87 & 0.13 & 2023-05-19T18:19:11 & 1.222 & 0.95 & 227.7 & 3494.3 & 4463 \\
    GECKO23bm & 105.223 & -24.254 & 21.41 & 0.17 & 2023-05-19T18:19:11 & 1.222 & 0.95 & 227.7 & 4172.0 & 4463 \\
    GECKO23bn & 104.981 & -24.151 & 20.02 & 0.07 & 2023-05-19T18:19:11 & 1.222 & 0.95 & 227.7 & 3317.9 & 4463 \\
    GECKO23bo & 105.035 & -24.034 & 20.72 & 0.11 & 2023-05-19T18:19:11 & 1.222 & 0.95 & 227.7 & 3452.4 & 4463 \\
    GECKO23bp & 104.944 & -23.905 & 20.76 & 0.11 & 2023-05-19T18:19:11 & 1.222 & 0.95 & 227.7 & 3167.3 & 4463 \\
    GECKO23bq & 104.959 & -23.848 & 20.22 & 0.10 & 2023-05-19T18:19:11 & 1.222 & 0.95 & 227.7 & 3243.2 & 4463 \\
    GECKO23br & 105.058 & -23.821 & 20.25 & 0.10 & 2023-05-19T18:19:11 & 1.222 & 0.95 & 227.7 & 3581.6 & 4463 \\
    GECKO23bs & 104.486 & -24.204 & 21.09 & 0.16 & 2023-05-19T18:19:11 & 1.222 & 0.95 & 227.7 & 1808.7 & 4463 \\
    GECKO23bt & 103.942 & -25.937 & 20.79 & 0.11 & 2023-05-19T18:19:11 & 1.222 & 1.0 & 301.9 & 270.9 & 10091 \\
    GECKO23bu & 103.901 & -25.752 & 21.08 & 0.13 & 2023-05-19T18:19:11 & 1.222 & 1.0 & 160.2 & 204.4 & 10964 \\
    \textbf{GECKO23bv} & 103.700 & -25.718 & 20.75 & 0.13 & 2023-05-19T18:19:11 & 1.222 & 0.99 & 145.3 & 69.9 & 9452 \\
    GECKO23bw & 101.756 & -22.653 & 21.38 & 0.16 & 2023-05-19T17:31:12 & 1.189 & 0.95 & 286.1 & 121.6 & 4289 \\
    GECKO23bx & 101.850 & -22.649 & 21.46 & 0.14 & 2023-05-19T17:31:12 & 1.189 & 0.99 & 161.0 & 253.3 & 8602 \\
    GECKO23by & 101.874 & -22.647 & 21.06 & 0.14 & 2023-05-19T17:31:12 & 1.189 & 0.99 & 161.0 & 265.3 & 8602 \\
    GECKO23bz & 101.866 & -22.608 & 21.49 & 0.16 & 2023-05-19T17:31:12 & 1.189 & 0.99 & 301.4 & 264.8 & 8594 \\
    GECKO23ca & 101.837 & -21.857 & 20.57 & 0.09 & 2023-05-19T17:31:12 & 1.189 & 0.95 & 233.1 & 193.6 & 5818 \\
    GECKO23cb & 101.748 & -21.808 & 20.54 & 0.06 & 2023-05-19T17:31:12 & 1.189 & 1.0 & 111.2 & 58.5 & 13688 \\
    GECKO23cc & 101.018 & -22.857 & 21.99 & 0.13 & 2023-05-19T17:31:12 & 1.189 & 0.9 & 291.4 & 261.6 & 2699 \\
    GECKO23cd & 100.671 & -19.893 & 19.99 & 0.05 & 2023-05-19T17:50:06 & 1.202 & 0.5 & 230.1 & 318.2 & 298 \\
    GECKO23ce & 99.833 & -22.008 & 20.42 & 0.06 & 2023-05-19T17:50:06 & 1.202 & 0.95 & 221.3 & 366.5 & 4850 \\
    GECKO23cf & 99.397 & -20.024 & 21.21 & 0.09 & 2023-05-19T17:50:06 & 1.202 & 0.95 & 294.7 & 255.1 & 5936 \\
    GECKO23cg & 101.529 & -19.047 & 20.93 & 0.08 & 2023-05-19T17:10:55 & 1.175 & 0.9 & 256.4 & 764.0 & 1625 \\
    GECKO23ch & 101.490 & -19.012 & 21.26 & 0.13 & 2023-05-19T17:10:55 & 1.175 & 0.9 & 270.5 & 857.1 & 1779 \\
    GECKO23ci & 101.513 & -18.913 & 20.21 & 0.09 & 2023-05-19T17:10:55 & 1.175 & 0.99 & 147.2 & 614.7 & 8106 \\
    GECKO23cj & 101.519 & -18.901 & 20.08 & 0.07 & 2023-05-19T17:10:55 & 1.175 & 0.99 & 147.2 & 573.3 & 8106 \\
    GECKO23ck & 101.573 & -18.812 & 21.09 & 0.15 & 2023-05-19T17:10:55 & 1.175 & 0.99 & 147.2 & 273.9 & 8106 \\
    GECKO23cl & 101.481 & -18.232 & 16.11 & 0.00 & 2023-05-19T17:10:55 & 1.175 & 0.9 & 254.4 & 1402.8 & 3564 \\
    GECKO23cm & 101.378 & -18.079 & 20.61 & 0.06 & 2023-05-19T17:10:55 & 1.175 & 0.9 & 254.4 & 1747.7 & 3564 \\
    GECKO23cn & 101.537 & -18.070 & 20.68 & 0.11 & 2023-05-19T17:10:55 & 1.175 & 0.9 & 254.4 & 1995.2 & 3564 \\
    GECKO23co & 101.582 & -18.035 & 20.34 & 0.10 & 2023-05-19T17:10:55 & 1.175 & 0.9 & 254.4 & 2182.6 & 3564 \\
    GECKO23cp & 95.637 & -11.426 & 19.99 & 0.10 & 2023-05-19T17:15:29 & 1.178 & 0.9 & 289.0 & 1365.1 & 3736 \\
    GECKO23cq & 95.717 & -10.585 & 19.77 & 0.09 & 2023-05-19T17:15:29 & 1.178 & 0.9 & 136.6 & 446.5 & 3245 \\
    GECKO23cr & 116.304 & -67.010 & 20.52 & 0.06 & 2023-05-19T01:11:36 & 0.509 & 0.9 & 293.0 & 597.9 & 3031 \\
    GECKO23cs & 115.357 & -66.780 & 19.32 & 0.02 & 2023-05-19T01:11:36 & 0.509 & 0.95 & 231.1 & 363.9 & 4263 \\
    GECKO23ct & 124.984 & -69.945 & 21.28 & 0.09 & 2023-05-19T01:04:32 & 0.504 & 0.9 & 231.3 & 4257.8 & 2584 \\
    GECKO23cu & 122.530 & -69.217 & 21.48 & 0.17 & 2023-05-19T01:04:32 & 0.504 & 1.0 & 217.6 & 2043.9 & 11189 \\
    GECKO23cv & 124.276 & -72.757 & 22.27 & 0.17 & 2023-05-19T00:22:42 & 0.475 & 0.99 & 233.1 & 743.5 & 6350 \\
    GECKO23cw & 121.506 & -72.854 & 20.59 & 0.06 & 2023-05-19T00:22:42 & 0.475 & 0.99 & 293.1 & 494.0 & 8426 \\
    GECKO23cx & 132.421 & -73.701 & 20.77 & 0.11 & 2023-05-19T01:34:09 & 0.524 & 1.0 & 205.9 & 7182.8 & 12744 \\
    GECKO23cz & 130.054 & -73.593 & 21.26 & 0.12 & 2023-05-19T01:34:09 & 0.524 & 1.0 & 205.9 & 5096.0 & 12744 \\
    GECKO23da & 130.851 & -73.217 & 21.00 & 0.10 & 2023-05-19T01:34:09 & 0.524 & 1.0 & 205.9 & 6495.6 & 12744 \\
    GECKO23db & 127.253 & -74.740 & 18.51 & 0.01 & 2023-05-19T01:34:09 & 0.524 & 1.0 & 205.9 & 2675.6 & 12744 \\
    GECKO23dc & 127.560 & -74.673 & 21.79 & 0.13 & 2023-05-19T01:34:09 & 0.524 & 1.0 & 205.9 & 2720.3 & 12744 \\
    GECKO23dd & 127.435 & -74.279 & 21.63 & 0.11 & 2023-05-19T01:34:09 & 0.524 & 1.0 & 205.9 & 2005.2 & 12744 \\
    GECKO23de & 127.290 & -74.170 & 18.64 & 0.02 & 2023-05-19T01:34:09 & 0.524 & 1.0 & 205.9 & 1840.4 & 12744 \\
    GECKO23df & 125.776 & -74.681 & 21.77 & 0.10 & 2023-05-19T01:34:09 & 0.524 & 0.99 & 187.7 & 1720.7 & 8225 \\
    GECKO23dg & 127.912 & -76.191 & 18.43 & 0.04 & 2023-05-19T01:24:51 & 0.518 & 1.0 & 208.1 & 2708.2 & 11291 \\
    GECKO23dh & 113.161 & -52.317 & 19.87 & 0.05 & 2023-05-18T23:47:45 & 0.450 & 0.99 & 181.9 & 91.3 & 7499 \\
    GECKO23di & 113.618 & -62.449 & 22.01 & 0.14 & 2023-05-19T18:57:44 & 1.249 & 0.99 & 285.9 & 172.3 & 8541 \\
    GECKO23dj & 113.464 & -62.400 & 21.78 & 0.15 & 2023-05-19T18:57:44 & 1.249 & 0.99 & 285.9 & 189.1 & 8541 \\
    GECKO23dk & 117.808 & -70.049 & 22.17 & 0.16 & 2023-05-18T18:17:40 & 0.221 & 0.99 & 310.9 & 236.8 & 9389 \\
    GECKO23dl & 117.339 & -69.400 & 20.70 & 0.09 & 2023-05-18T18:17:40 & 0.221 & 0.99 & 298.4 & 642.8 & 6885 \\
    GECKO23dm & 115.045 & -71.538 & 18.31 & 0.01 & 2023-05-18T18:05:23 & 0.213 & 0.5 & 170.8 & 17.8 & 118 \\
    GECKO23dn & 142.658 & -81.890 & 21.69 & 0.09 & 2023-05-19T19:32:50 & 1.273 & 1.0 & 202.1 & 4969.6 & 12448 \\
    GECKO23do & 138.229 & -81.939 & 21.31 & 0.10 & 2023-05-19T19:32:50 & 1.273 & 1.0 & 202.1 & 2998.6 & 12448 \\
    GECKO23dp & 138.762 & -81.325 & 21.08 & 0.08 & 2023-05-19T19:32:50 & 1.273 & 1.0 & 223.1 & 4573.8 & 11734 \\
    GECKO23dq & 138.900 & -81.272 & 20.84 & 0.08 & 2023-05-19T19:32:50 & 1.273 & 1.0 & 223.1 & 4713.5 & 11734 \\
    GECKO23dr & 139.238 & -81.207 & 20.75 & 0.08 & 2023-05-19T19:32:50 & 1.273 & 1.0 & 223.1 & 4978.7 & 11734 \\
    GECKO23ds & 139.496 & -81.188 & 21.31 & 0.07 & 2023-05-19T19:32:50 & 1.273 & 1.0 & 223.1 & 5136.0 & 11734 \\
    GECKO23dt & 139.157 & -81.045 & 20.81 & 0.09 & 2023-05-19T19:32:50 & 1.273 & 1.0 & 226.4 & 5168.7 & 12013 \\
    GECKO23du & 138.980 & -80.836 & 20.75 & 0.08 & 2023-05-19T19:32:50 & 1.273 & 1.0 & 226.4 & 4902.7 & 12013 \\
    GECKO23dv & 139.145 & -80.827 & 21.11 & 0.11 & 2023-05-19T19:32:50 & 1.273 & 1.0 & 226.4 & 4991.9 & 12013 \\
    GECKO23dw & 132.423 & -83.011 & 21.58 & 0.07 & 2023-05-19T19:32:50 & 1.273 & 1.0 & 142.6 & 568.0 & 11856 \\
    GECKO23dx & 105.228 & -29.302 & 19.95 & 0.08 & 2023-05-19T18:15:24 & 1.220 & 0.99 & 277.4 & 27.1 & 5982 \\
    \textbf{GECKO23dz} & 115.525 & -74.705 & 22.18 & 0.14 & 2023-05-19T09:19:01 & 0.847 & 0.5 & 230.4 & 15.7 & 449 
    \label{tab:candidate}
\end{longtable*}

\appendix

\acknowledgments
This work was supported by the National Research Foundation of Korea (NRF) grants, No. 2020R1A2C3011091, and No. 2021M3F7A1084525, funded by the Korea government (MSIT).
SWC acknowledges the support from the Basic Science Research Program through the NRF funded by the Ministry of Education (RS-2023-00245013). 
We thank the operators at KMTNet for performing the requested observations.
This research has made use of the KMTNet system operated by the Korea Astronomy and Space Science Institute (KASI) at three host sites CTIO in Chile, SAAO in South Africa, and SSO in Australia. Data transfer from the host site to KASI was supported by the Korea Research Environment Open NETwork (KREONET).
We thank iTelecope.Net and its staff Brad Moore, Pete Poulos, and Ian Leeder for their management and support for LSGT observation. We thank Jeeun Hwang for her assistance in obtaining Pan-STARRS DR1 snapshots of potential host galaxies.
The national facility capability for SkyMapper has been funded through ARC LIEF grant LE130100104 from the Australian Research Council, awarded to the University of Sydney, the Australian National University, Swinburne University of Technology, the University of Queensland, the University of Western Australia, the University of Melbourne, Curtin University of Technology, Monash University and the Australian Astronomical Observatory. SkyMapper is owned and operated by The Australian National University's Research School of Astronomy and Astrophysics. The survey data were processed and provided by the SkyMapper Team at ANU. The SkyMapper node of the All-Sky Virtual Observatory (ASVO) is hosted at the National Computational Infrastructure (NCI). Development and support of the SkyMapper node of the ASVO has been funded in part by Astronomy Australia Limited (AAL) and the Australian Government through the Commonwealth's Education Investment Fund (EIF) and National Collaborative Research Infrastructure Strategy (NCRIS), particularly the National eResearch Collaboration Tools and Resources (NeCTAR) and the Australian National Data Service Projects (ANDS).

\vspace{5mm}
\facilities{KMTNet(1.6-m; SAAO, SSO, CTIO), RASA36(0.35-m), LSGT(0.43-m)}


\software{
        astropy \citep{2013A&A...558A..33A,2018AJ....156..123A,2022ApJ...935..167A}
        astroquery \citep{2019AJ....157...98G},
        Astrometry.net \citep{2010AJ....139.1782L},
        ccdproc \citep{2015ascl.soft10007C},
        HOTPANTS \citep{2015ascl.soft04004B}, 
        SExtractor \citep{1996A&AS..117..393B},
        SCAMP \citep{2010ascl.soft10063B},
        }






\newpage
\bibliography{sample63}{}
\bibliographystyle{aasjournal}

\end{document}